\documentclass{JHEP}

\usepackage{rotating} 
\usepackage{rotating} 
\newcommand{\be}{\begin{equation}}
\newcommand{\ee}{\end{equation}}

\newcommand{\bea}{\begin{eqnarray}}
\newcommand{\eea}{\end{eqnarray}}
\newcommand{\bean}{\begin{eqnarray*}}
\newcommand{\eean}{\end{eqnarray*}}

\def\beq{\begin{equation}}

\def\eeq{\end{equation}}

\def\Tr{\mathop{\rm Tr}}
\def\diag{\mathop{\rm diag}}

\newcommand{\CN}{{\cal N}}
\newcommand{\CQ}{{\cal Q}}

\preprint{UPR-1032-T,  MIFP-03-05, {\tt hep-th/0303065}}

\title{Phases of $\mathbf{\CN=1}$ Supersymmetric Gauge Theories with Flavors}
\author{
Vijay Balasubramanian$^1$,
Bo Feng$^2$,  
Min-xin Huang$^1$, and Asad Naqvi$^1$ \\
~\\
\[
\begin{tabular}{ll}
\begin{tabular}{l}
1. David Rittenhouse Laboratories,\\
University of Pennsylvania,\\
209  S.~33rd St.,
Philadelphia, PA 19104-6396\\
\end{tabular}
&
\begin{tabular}{l}
2. Institute for Advanced Study, \\
Olden Lane, Princeton, NJ 08540 \\
\\
\end{tabular}
\end{tabular}\]
\email{vijay@physics.upenn.edu, fengb@ias.edu,}
\email{minxin@sas.upenn.edu, naqvi@physics.upenn.edu}
}

\abstract{
We explore the phases of $\CN = 1$ supersymmetric $U(N)$ gauge theories with fundamental matter that arise as deformations of $\CN = 2$ SQCD by the addition of a superpotential for the adjoint chiral multiplet.  As the parameters in the superpotential are varied, the vacua of this theory sweep out  various branches, which in some cases have multiple semiclassical limits. In such limits, we recover
the vacua of various product gauge group theories, with flavors charged under some group factors. 
We describe in detail the structure of  the vacua  in both classical and quantum regimes, and 
develop general techniques such as an addition and a multiplication map which relate vacua of different gauge theories. We also consider possible indices characterizing different branches and potential relationships with matrix models. 
}

\begin{document}
\section{Introduction}

Traditionally, there are two senses in which two quantum field theories can be ``dual" to each other.   On the one hand, two theories can be {\it equivalent} to each other in the sense that all the correlation functions of one can be computed from the other (and vice versa) by a suitable identification of dual variables.  An example is the electric-magnetic duality of the $\CN = 4$ super Yang-Mills theory.   Another kind of duality, described by Seiberg for $\CN = 1$ supersymmetric field theories 
occurs when two different microscropic theories have identical macroscopic (or infra-red) dynamics.   For example, an $\CN = 1$ supersymmetric  $SU(N_c)$ gauge theory with $3N_c > N_f > N_c+1$ fundamental flavours has the same long distance physics as an $SU(N_f - N_c)$ gauge theory with $N_f$ flavours.
Recently, Cachazo, Seiberg and Witten have proposed another notion of duality in certain $\CN = 1$  supersymmetric field theories \cite{Cachazo:2002zk}.   Related observations had been made earlier by Friedmann \cite{tamar} in the context of G2 holonomy compactifications and also more recently  by Ferrari \cite{ferrari}.

This new notion of duality has arisen in the context of
the $\CN=1$ $U(N)$ theory with an adjoint chiral multiplet $\Phi$ and superpotential $W(\Phi)$.  This theory has the same matter content as the $\CN=2$ pure $U(N)$ gauge theory, and can be thought of as a deformation of that theory by the addition of a superpotential $\Tr W(\Phi)$. The fascinating relation of this theory to a bosonic matrix model was uncovered by Dijkgraaf and Vafa in \cite{dv1,dv2,dv3}. It was shown that the superpotential for the glueball superfield can be computed in an auxiliary matrix model of a single $M \times M$ matrix $\hat{\Phi}$ with potential $ \Tr W(\hat{\Phi})$ in the $M\rightarrow \infty$ limit.

At the classical level, the theory has vacua in which the diagonal vacuum expectation value for $\Phi$ breaks the gauge group down to $\prod_i U(N_i)$. The VEV for $\Phi$ has diagonal entries $a_i$, each repeated $N_i$ times, where $a_i$ are roots of the polynomial $W'(x)$. Quantum mechanically, the classical vacua lead to $\prod_i N_i$ different vacua. This description in terms of the unbroken product gauge group only makes sense if the difference between $a_i$'s is much larger than $\Lambda$, the dynamical scale of the $\CN=2$ theory. It is interesting to vary $a_i$ by varying the parameters in the superpotential to pass through a region where the $a_i$ are approximately equal to each other, before being far apart once again. In such a situation, it is possible that a vacuum of the $\prod_i U(N_i)$ theory is continuously transformed into a vacuum of $\prod_i U(\widetilde{N}_i)$ gauge theory, where $\widetilde{N}_i$ are not the same as $N_i$. 
Such a phenomenon has been identified and studied in the context of G2 holonomy compactifications of M-theory \cite{tamar} and in the context of pure SUSY gauge theory
\cite{Cachazo:2002zk}.   
The possibility of a smooth transition of this kind between two microscopically different theories exists only in cases where the theories in question possess vacua with the same low energy dynamics, since it is not then necessary to encounter any singular loci such as points of enhanced symmetry during the interpolation.
It has been proposed that such smooth transitions should be understood as a new kind of duality, relating microscopic theories which have vacua with identical low energy dynamics \cite{tamar, Cachazo:2002zk}.

It is interesting to understand how these ideas extend to the case where fundamental matter is included.\footnote{The generalization of \cite{Cachazo:2002zk}
to some other cases has been made in \cite{Ahn:2003cq,oz}}   In particular, how does Seiberg duality of $\CN = 1$ theories fit into the story of \cite{Cachazo:2002zk}?
Accordingly,
we study $\CN=1$ $U(N)$ theory with an adjoint chiral multiplet $\Phi$, and $N_f$ fundamental and anti-fundamental chiral multiplets $\CQ_i$ and $\tilde{\CQ^i}$ ($i =1 \dots N_f$). We also add a superpotential $W= W(\Phi)+ \sqrt{2} \widetilde{\CQ}_i \Phi \CQ^i+
\sqrt{2}m_i \widetilde{\CQ}_i  \CQ^i $. This model is an $\CN=2$ $U(N)$ gauge theory with $N_f$ fundamental hypermultiplets which is deformed to $\CN=1$ by addition of the tree level superpotential $\Tr W(\Phi)$. This theory has classical vacua in which the gauge group is broken down to $\prod_{i=1}^n U(N_i)$, with $M_i$ fundamental flavors charged under the $U(N_i)$ factor.  As in the case without flavors, this description in terms of product gauge groups is only valid in the regime of parameters in the superpotential when the roots $a_i$ of $W'(x)$ are widely separated ($|a_i-a_j| \gg \Lambda$) and the gauge group $U(N)$ is broken down to $\prod_i U(N_i)$ at a scale at which the $U(N)$ gauge coupling is weak.  The quantum dynamics in each of these factors depends critically  on $M_i$: an Affleck, Dine, Seiberg (ADS) superpotential is generated for $M_i < N_i$;
confinement
with a quantum deformation of the moduli space
occurs when $M_i=N_i$; confinement 
without quantum corrections to the moduli space occurs when $M_i=N_i+1$;  a  magnetic description in term of dual quarks and 
gluons is relevant when $M_i > N_i + 1$,
leading to a dual gauge symmetry $U(M_i-N_i)$ with $M_i$ fundamental quarks and a meson singlet $M$. By incorporating these phenomena in our analysis for each gauge group factor, we can determine the vacua of the product gauge theory. 

We can then ask a similar  question in the presence of flavors to the one asked in \cite{Cachazo:2002zk}:  by varying the parameters in the superpotential, can we follow these `weak coupling vacua' into the `strong coupling region', before taking another limit in the space of parameters which takes us to perhaps another weak coupling region? In fact, we will show that  it is possible, by a smooth variation of parameters in the superpotential,  to continuously interpolate between vacua of theories with different underlying product gauge groups and different arrangements of charged fundamental matter under each gauge group factor. 
This leads to a picture where, as we modifiy the tree-level superpotential $\Tr W(\Phi)$, a low energy observer will see the same physics while the natural underlying microscopic dynamical variables change completely.
Continuous interpolation is not possible between {\it all} vacua of a theory with the same low energy behaviour -- most interpolations will pass through singular points in the moduli space.
Indeed, we point out  a set of selection rules that restrict which classical vacua admit continuous interpolations between each other.  More accurately, these selection rules provide necessary but not sufficient conditions  to establish which vacua of classical theories we can interpolate between.

In Sec. \ref{classical}, we review the classical moduli space of $\CN=2$ SQCD deformed to $\CN=1$ by adding a superpotential $\Tr W(\Phi)$. The pure Coulomb branch (where all squarks have vanishing VEVs) only exists if $W'(-m_i)=0$ where $m_i$ are the quark masses. We will mostly restrict ourselves to superpotentials which satisfy this condition. On the pure Coulomb branch, the adjoint acquires a diagonal VEV, with the diagonal entries being roots $a_i$ of the polynomial $W'(x)$. With $N_i$ of the diagonal entries being $a_i$, the gauge group is broken down to $\prod_i U(N_i)$, with $M_i$ flavors charged under the $U(N_i)$ factor.  For each of these $U(N_i)$ factors, there is a rich structure of Higgs branches, indexed by an integer $r_i$. 
These Higgs branches meet the Coulomb branch along a submanifold called the root of the $r$-th Higgs branch. 

In Sec. \ref{quantum}, we explore the quantum theory. We employ two approaches which are valid in different regimes of parameters in  the superpotential. We first consider the case when the difference between the roots of $W'(x)$ is much greater than $\Lambda$, the $\CN=2$ dynamical scale. In this regime (the weak coupling region), the semiclassical description in term of product gauge groups is valid and we can analyze each gauge group factor separately.  The adjoint field with mass $\mu \gg \Lambda$ can be integrated out from the low energy theory.  At scales below $\Lambda_1$ (given by the matching relation $\Lambda_1^{3N_i-M_i}=\mu^{N_i} \Lambda^{2N_i-M_i}$), the strong coupling dynamics of the resulting $U(N_i)$ theory with $M_i$ flavors (which depends critically on $M_i$)  becomes relevant. We analyze the vacuum structure of these $\CN=1$ theories in detail. In these vacua, the meson $M=Q\tilde{Q}$ has a non-zero VEV, which can be characterized by an integer $r$. In addition, if $M_i \geq N_i$, there are vacua in which the  baryons and dual quarks  have non-zero VEVs. 

When  roots of $W'(x)$ are nearly the same, it is more appropriate to consider the $\CN=1$ theory as a small perturbation of a strongly coupled $\CN=2$ gauge theory with no superpotential.  The description of the theory in terms of product gauge groups is not appropriate. To find the vacua in this regime, we start by characterizing the submanifold of the $\CN=2$ Coulomb branch which is not lifted by the $\CN=1$ deformation.  This  consists of special points in the roots of the various $r$-th Higgs branches where a certain number of mutually local magnetic monopoles become massless.  We will label our branches with index $r$. Furthermore,  it turns out 
we must distinguish between two types of branches -- the ``baryonic" and ``non-baryonic" branches, which have different numbers of massless magnetic monopoles.   The baryonic branches only exists for certain special values of $r$.  
On  each branch, the Seiberg-Witten curve, which encodes the quantum dynamics of the  $\CN=2$ theory, factorizes in a particular way. 
Extremizing the superpotential on this factorization locus gives us the $\CN=1$ vacua. By a holomorphic variation of the parameters in the superpotential, we can take a semiclassical limit which transports these vacua to a regime in which they are appropriately interpreted as vacua of a product gauge theory with charged flavors under the group factors. These are the vacua  discussed in the previous paragraph. In fact, occasionally, it will be possible to take multiple classical limits such that the same vacuum in the strong coupling region can be transported to multiple weak coupling regions where it becomes the  vaccum of different product group theories. 

In Sec. \ref{maps}, we  develop general techniques which we  use to relate vacua of different theories. In particular, we introduce an {\it addition map}, which  relates vacua of a $U(N)$ theory with $N_f$ flavors in the $r$-th branch to vacua of a $U(N+d)$ theory, with $N_f+2d$ flavors on the $r+d$ branch. This map proves to be very useful  in later sections. We also generalize the muliplication map of \cite{Cachazo:2001jy,Cachazo:2002zk} to include flavors. This can be used to relate vacua of a $U(N)$ theory to vacua of a $U(tN)$ theory.  The multiplication map suggests how we might
define a new index in the presence of flavors that is in analogy with the ``confinement index'' defined in the flavorless case  case of \cite{Cachazo:2002zk} which would  enable us to distinguish different branches of vacua.

In Sec. \ref{sec:quad}, we discuss the phase structure of our theory when the superpotential is quadratic in the adjoint field. We only consider cases with all the flavors having the same mass. 
For a $U(N)$ theory with $N_f$ flavors, we analyze two special branches (with $r=N_f/2$ and the ``baryonic" $r=N_f-N$) where a general discussion is possible. Otherwise,  
we mostly restrict ourselves to working out specific examples. In particular, we discuss theories with  $U(2)$ and $U(3)$ gauge groups and an even number of flavors. We find the vacua in the strong coupling region on the various branches and transport them to the weak coupling regime by variation of the parameters in the superpotential. We check that the number of strong coupling vacua on a particular branch matches with the number of vacua in the weak coupling region.  In the case of a quadratic superpotential no interpolations between different classical vacua are possible, but the phase structure is nevertheless interesting.

In Sec. \ref{sec:cubic}, we consider the theory with a cubic superpotential and analyze in detail the  phase structure of specific theories on  various branches. In particular, we consider the $U(2)$, $U(3)$ and $U(4)$ theories with various number of flavors. In the semiclassical limits, the we approach a theory with two gauge group factors and flavors charged under one of the factors.  We analyze the classical limits of various branches, count the number of
vacua  and find  a match between the weak and strong coupling analyses. In particular, on some branches of $U(4)$ theory with zero, two and four flavors, 
we find smooth interpolations amongst different classical vacua.  The phase structure is different for different numbers of flavors. 
When no flavors are present, a smooth interpolation exists  between some vacua of $U(2) \times U(2)$ and those of $U(1) \times U(3)$ \cite{Cachazo:2002zk}.  With two flavors, the three  classical limits, {\em viz} , $U(4)\rightarrow U(2)\times U(2)$ with flavors charged under one $U(2)$ factor,
$U(4)\rightarrow U(1)\times U(3)$ with flavors charged under $U(1)$, and $U(4)\rightarrow U(1)\times U(3)$ with flavors charged under the  $U(3)$ factor,  are all smoothly connected. However,  with four flavors,  the classical limit in which  flavors are charged under  $U(2)$ can only be connected with  the one in which  they are charged under $U(3)$.  In addition, there is a different branch on which  the $U(1) \times U(3)$ theories with flavors charged under $U(1)$ and $U(3)$ are smoothly connected.  
Recall that a $U(3)$ theory with four flavors has a dual description in terms of a $U(1)$ theory with four flavors. Thus the smooth interpolation on this branch relates vacua of two Seiberg dual theories.\footnote{This is slightly  inaccurate. The two theories in the dual pair are not precisely related by Seiberg duality, but are deformations, by a relevant operator, of theories that are Seiberg dual. }
With six flavors, the phase structure of $U(4)$ theory is quite different: there is no smooth interpolation between vacua of the three classical limits mentioned above.

We end this section by briefly mentioning some assumptions made in this paper to simplify analysis: 

\begin{enumerate}
\item All the flavors have the same mass $m_i=m$. The generalization to the case with different masses is straightforward. As described above the branches of a single gauge group factor with massless fundamentals are labeled by an integer $r$.  When there are different masses, the possible branches will be more intricate and will be labeled by a set of integers  $(r_1,r_2,...,r_k)$.   More interesting interpolations between vacua can then occur.

\item
We have restricted ourselves  to superpotentials $W(x)$ such that $-m$  is one of the roots of $W'(x)$. This is for the following reason: if $W'(-m)\neq 0$, $Q,\widetilde{Q}$ (which are flavors under the gauge group factor $U(N_i)$ arising from $N_i$ diagonal entries of $-m$ in the VEV of $\Phi$)  must necessarily have non-zero VEVs and hence the gauge group $U(N_i)$ would be completely Higgsed. Thus we will be reduced to the case in which there are no massless flavors under any of the gauge group factor. The latter situation has already been discsussed in \cite{Cachazo:2002zk}.  

\item We mostly consider an even number of flavors (except for one example). The reason is that the computations in examples with an odd number of flavors are more involved. However, we expect that the difference between even and odd flavors is technical and will not lead to significantly different physical phenomena. 
\end{enumerate}

\section{The Classical Moduli Space}
\label{classical}
We will be studying  $\CN=2$ supersymmetric  $U(N)$ gauge theory 
with $N_f$ fundamental  hypermultiplets, with the supersymmetry broken down
to $\CN=1$ by addition of a tree level superpotential for  $\Phi$,
the adjoint chiral superfield in the $\CN=2$ gauge multiplet:
\be \label{superpotential-W}
W= W(\Phi)+ \sqrt{2} \widetilde{\CQ}^a_i \Phi^b_a \CQ_b^i+
\sqrt{2} \widetilde{\CQ}^a_i m^i_j \CQ^j_a,
\ee
where 
\be \label{W-Phi}
W(\Phi)=\sum_{j=1}^{p+1} g_j u_j =\sum_{j=1}^{p+1} g_j {\Tr(\Phi^j) \over j}.
\ee
The mass matrix is diagonal as
\be \label{mass}
m={\mathrm {diag}}[ m_1 I_{M_1},m_2 I_{M_2}, ...,m_k I_{M_k}], ~~~~\sum_{i=1}^k M_i=N_f, 
\ee
with different $m_i$.
The anomaly free global symmetry is $\prod_{a=1}^k SU(M_a)$.

An important property of this theory is the existence of classical flat directions.\footnote{ The moduli space of this theory but with gauge group $SU(N)$ instead of $U(N)$ has been discussed extensively in the literature
 \cite{Argyres:1996eh,Hori:1997ab,deBoer:1997ap,
Carlino:2000uk}. We will adapt these results for the $U(N)$ theory.} These are determined by solving the F and D-term equations:
\bea
0 & = & [\Phi,\Phi^\dagger]  \label{adjoint-D} \\
0 & = & \CQ^i_a (\CQ^\dagger)_i^b - (\widetilde{\CQ}^\dagger)^i_a 
\widetilde{\CQ}_i^b ~~~~~~~~~{\mathrm{D-terms}}\label{flavor-D}\\
&& \nonumber \\
0 & = & W'(\Phi)^a_b+\sqrt{2} \widetilde{\CQ}^a_i  \CQ_b^i  \label{F-Phi} \\
0 & = & \widetilde{\CQ}^a_j \Phi^b_a + \widetilde{\CQ}^b_i m^i_j \label{wide-Q}~~~~~~~~~~~~~~~{\mathrm {F-terms}}\\
0 & = &  \Phi^b_a \CQ_b^i+m^i_j \CQ^j_a \label{Q} 
\eea
Eq (\ref{adjoint-D}) implies that $\Phi$ can be brought into a diagonal form via a gauge transformation: \be  \label{Phi-form}
\Phi=[\lambda_1 I_{N_1}, \lambda_2 I_{N_2},...., \lambda_s 
I_{N_s}],~~~
\sum_{i=1}^s N_i=N.
\ee
The gauge symmetry is broken down to $U(N)\rightarrow 
\prod_{i=1}^s U(N_i)$. 

$\CQ_a^i, ({}^t\widetilde{\CQ})_a^i$ are $N \times N_f$ matrices. From 
eq (\ref{wide-Q}) and (\ref{Q}), we can see that these matrices are  non-zero if and only if $\lambda_i+m_j= 0$ for some $i,j$.  This simply reflects the statement that the Higgs branch emanates from the point in the Coulomb branch where the quarks are massless. By relabeling indices, we can choose $\lambda_u=-m_u$ for $u = 1, \dots, h$. Then, $\CQ$ and $\tilde{\CQ}$ are:
\be  \label{Q-block}
\CQ=\left[  \begin{array}{cccc}  
Q_1 & 0 & 0 & 0 \\   0 & Q_2 & 0 & 0 \\ 0 & 0 & Q_h & 0 \\
0 & 0 & 0 & 0  
\end{array}  \right],~~~~~~^t\widetilde{Q}=\left[  \begin{array}{cccc}  
~^t\widetilde{Q}_1 & 0 & 0 & 0 \\   
0 & ~^t\widetilde{Q}_2 & 0 & 0 \\ 
0 & 0 & ~^t\widetilde{Q}_h & 0 \\
0 & 0 & 0 & 0  
\end{array}  \right],
\ee
where $Q_u, ~^t\widetilde{Q}_u~$ are $N_u\times M_u$ matrices. With this form for $\CQ$ and $\tilde{\CQ}$ and the diagonal form for $\Phi$ (eq (\ref{Phi-form})), the F-term equation, eqs (\ref{wide-Q}) and (\ref{Q})  become\footnote{Here, we abuse notation slightly. The indices $i$, $a$ and $b$ run over the colors and massless flavors in a given block $U(N_u)$. This is different from the role of these indices in  (\ref{flavor-D}) and (\ref{wide-Q}), where they represented color and flavor indices in the full $U(N)$.}
\bea
{-1\over \sqrt{2}} W'(-m_u) \delta_a^b& = 
& (Q_u)^i_a (\widetilde{Q}_u)^b_i,~~~~~~u=1,....,h \label{reduced-F-1} \\
W'(\lambda_v) & = & 0,~~~~v=h+1,...,k.  \label{reduced-F-2}
\eea
Furthermore, the D-terms in  (\ref{flavor-D}) decompose into D-terms for each
$u$'th block: 
\be \label{reduced-D}
0 = Q_u Q_u^\dagger - \widetilde{Q}^\dagger_u \widetilde{Q}_u,~~~~u=1,...,h
\ee
In this way, we obtain the reduced F (\ref{reduced-F-1}) and D-term (\ref{reduced-D}) conditions for each $u$'th block. The solutions to these conditions are indexed by an integer $r_u$ with 
\be \label{r-region}
r_u \leq {\mathrm{min}}({M_u \over 2} , N_u).
\ee
In particular, on the $r$th Higgs branch\footnote{We will drop the index $u$ to simplify notation.}, 
\begin{eqnarray}  
Q&=&\left[ \begin{array}{lll}  
K_{r\times r}  &  0_{r\times r}  & 0_{r\times (M_u-2r)} \\
0_{(N_u-r)\times r} & 0_{(N_u-r)\times r} & 0_{(N_u-r)\times (M_u-2r)} \\
\end{array} \right], \nonumber \\ && \nonumber \\
^t\widetilde{Q}&=&\left[ \begin{array}{lll}  
 P_{r\times r} & \widetilde{K}_{r\times r}  & 0_{r\times (M_u-2r)} \\
0_{(N_u-r)\times r} & 0_{(N_u-r)\times r} & 0_{(N_u-r)\times (M_u-2r)} \\
\end{array} \right],
\label{r-branch}
\end{eqnarray}
where $K$ is a diagonal $r\times r$ matrix with non-zero diagonal elements. 
If $W'(-m_u) \neq 0$, we must have $r=N_u$ and $P$ is a non-degenerate
 $r\times r$ matrix. Thus at a generic point on this branch, the gauge group factor $U(N_u)$ is completely
broken.   On the other hand, if $W'(-m_u)=0$, the solution has $P=0$ and 
$\widetilde{K}=K$.
Generically, on this branch, the unbroken gauge group is $U(N_u-r)$  
with $M_u-2r$ massless flavors and $r(M_u-r)$ neutral hypermultiplets.
It is useful to compute the gauge invariant meson matrix $M_i^j$:
\be  \label{r-th-M}
M_i^j= \widetilde{Q}_i^a Q_a^j=\left[ \begin{array}{ccc}
(PK)_{r\times r}  & ~~(\widetilde{K}^t K)_{r\times r}~~ & 0_{r\times (M_u-2r)} \\
0_{r\times r}~~  & 0_{r\times r}~~ & 0_{r\times (M_u-2r)} \\
0_{{(M_u-2r)} \times r}  & 0_{(M_{u}-2r) \times r}~~ & 0_{{(M_u-2r)\times (M_u-2r)}} \\
\end{array} \right].
\ee
For the $r$'th branch, the meson has rank $r$.
The global flavor symmetry is spontaneously  broken down to $SU(M_u)\rightarrow SU(M_u-2r)\times U(1)^r$. 

\subsection*{\underline{Summary of the classical moduli space}}
We are now ready to discuss the complete picture of the classical moduli space. Above, we have described how the adjoint vevs break the gauge group $U(N)$ into a product of $U(N_i)$ factors, each with $M_i$ massless flavors. Below, we summarize the structure of the Coulomb and Higgs branches of each of these factors. 
\paragraph{Coulomb branch:}
On the Coulomb branch, $\CQ$ and $\tilde{\CQ}$ both vanish and the adjoint $\Phi$ has diagonal entries $\lambda_i$, where each  $\lambda_i$ is a root 
of the polynomial $W'(z)$ (\ref{reduced-F-2}). The
gauge group is broken to $U(N)\rightarrow 
\prod_{i=1}^s U(N_i)$. The Higgs branches (with $\CQ$ and $\tilde{\CQ}$ non-zero) touch the Coulomb branch at special points, called the roots of the Higgs branch. These points will only exist if $W'(x)$ has a zero at $x=-m_i$. This is clear from eq (\ref{reduced-F-2}): vanishing quark vevs necessarily imply that $W'(-m_i)=0$. 
\vspace{-0.3cm}
\paragraph{Higgs branches:} 
On the Higgs branch, the quarks generically have non-zero vevs. 
On such branches, the diagonal entries for the adjoint are $\lambda_i=-m_i$ if $Q_i$ and $\tilde{Q}_i$ are non-zero. $W'(-m_i)$ does not have to vanish. However, if  $W'(-m_i) \neq 0$, the gauge theory is necessarily Higgsed, and the $U(N_i)$ factor is completely broken. We will not pursue this case further in this paper. 

If $W'(-m_i)=0$, there is a richer structure of Higgs branches, indexed by an integer $r$. These $r$'th Higgs branch are shown in (\ref{r-branch}). At the root of the Higgs branch, $Q$ and $\tilde{Q}$ vanish and there are massless flavors charged under the group $U(N_i)$ at these points. In the next section, we will see how we can use the Seiberg-Witten curve describing the Coulomb branch of the $\CN=2$ theory (without the superpotential $W(\Phi)$) to study the dynamics at these roots. 
In order to explore these roots, 
we will need to tune the parameters of the theory (the masses $m_i$ and the parameters in the superpotential $W(\Phi)$) to  guarantee that $-m_i$ are roots of $W'(z)$ ( $W'(z)$ can have some
roots which are not equal to $-m_i$).

In summary, classically,  $U(N)$ with $N_f$ flavors is broken to $k$ non-interacting factors with
$U(N_i)$ gauge symmetry and $M_i$ massless flavors. Each factor has a Coulomb and various Higgs branches.  In the $r$'th Higgs branch, the unbroken gauge group is
$U(N_i-r)$  with $M_i-2r$ massless flavors and  $r(M_i-r)$ neutral hypermultiplets.

Below, we will see how this classical picture changes in the quantum theory and then we will explain how strong coupling dynamics enables one to interpolate smoothly between vacua of theories with different classical product gauge group and matter contents.

\section{The Quantum Theory}
\label{quantum}
The above classical picture is modified significantly in the quantum theory. The low energy dynamics on the Coulomb branch of the $\CN=2$ quantum theory  (without $W(\Phi)$) is described by the Seiberg-Witten curve. In such a theory, it was shown \cite{Argyres:1996eh} that the moduli space locally is a product of the Coulomb and Higgs branches (by using powerful non-renormalization theorems due to $\CN=2$ supersymmetry). The metric on the Higgs branch does not receive any quantum corrections. However, the location of the roots of the Higgs branch, {\em i.e.} the points where the Higgs branch meets the Coulomb branch will be modified in the quantum theory. On the other hand, if the theory has only  $\CN=1$ supersymmetry, the metric and even the topological structure of the Higgs branch is modified quantum mechanically. The familiar example of such a phenomenon is  SQCD  \cite{Seiberg:1994bz,Seiberg:1994pq}.

We can analyze the $\CN=1$ quantum theory in two ways. The first approach is to consider this theory as a small perturbation of a strongly coupled $\CN=2$ gauge theory with $W=0$.
We will call this the strong coupling analysis.
The Seiberg-Witten curve encodes the low energy quantum dynamics of the $\CN=2$ theory on the Coulomb branch. Turning on a tree-level superpotential lifts almost all points on the Coulomb branch, except points in the Higgs branch roots where a certain number of mutually local monopoles become massless. Furthermore, on this sub-manifold of the Coulomb branch, the tree-level superpotential has to be minimized to find the $\CN=1$ vacua.  By varying the parameters of the superpotential, these  $\CN=1$ vacua can be moved around on the $\CN = 2$ moduli space. In particular, special corners in the parameter space will place these vacua in regions where the $U(N)$ gauge symmetry breaking scale is much greater than $\Lambda$, the $\CN=2$ dynamical scale. Thus the gauge group is higgsed   when the gauge coupling is small and the description in terms of non-interacting product gauge group factors is valid. More concretely, we can integrate out (in each gauge group factor) the massive adjoint chiral field, $\Phi$, which has a mass well above $\Lambda$. The corresponding $\CN=1$ theory thus obtained, valid below scales of order $\mu$ can be analyzed in various group factors separately. In each factor, it will become strongly coupled in the infra-red  and will have vacua, details of which will depend on the number of flavors charged under the group factor. The vacua of the product group theory (which are just products of vacua of each factor)  thus obtained have been relocated from the strong gauge coupling region by variations of the parameters in the superpotential. 
We will call discussions of this region of moduli space the weak coupling analysis.

In summary, the strong coupling analysis involves treating the superpotential as a small perturbation around the $\CN=2$ theory, with all the non-trivial dynamics included in the SW curve. The weak coupling analysis involves integrating the adjoint out, and studying the non-trivial dynamics of the resulting $\CN=1$ theory. 
We can interpolate smoothly between vacua in the weak and strong coupling regions by holomorphic variation of the parameters in the superpotential $W(\Phi)$ because the theory has $\CN = 1$ supersymemtry.  Hence there is no phase transition in moving between these points.
What is perhaps most interesting is the fact that different weak coupling regions with different microscopic physics can be reached smoothly from the same strongly coupled point. Of course, this also provides a smooth interpolation between vacua of the two different $\CN=1$ theories obtained  in the weakly coupled region.

\subsection{\textbf{Weak gauge coupling analysis}}
\label{weak}
The weak coupling analysis is valid when the differences between the roots of the polynomial $W'(x)$ are much bigger than $\Lambda$.  In this case, we can trust the classical description of the theory in terms of low energy group $\prod_i U(N_i)$ with almost decoupled factors and adjoints $\Phi_i$ having a large mass of order $\mu \gg \Lambda$. The adjoints $\Phi_i$ can be integrated out from the low energy theory in each group factor separately. Therefore, we will focus our attention on a single $U(N_i)$ factor, with $M_i$ flavors.  In this integrating out procedure, we will only consider the quadratic part of the effective potential since the higher powers will be suppressed by $\mu$.   Hence 
our analysis below of the structure of the quantum mechanical vacua will only receive small perturbative corrections.  The relevant part of the superpotential is then
\be \label{w-mass}
W={\mu } \Tr(\Phi^2) + \sqrt{2} \widetilde{Q}^a_i \Phi^b_a Q_{b}^i.
\ee 
Integrating $\Phi$ out yields
\be \label{integrate-Phi}
W=-{1\over 2\mu} \Tr(M^2),~~~~M_i^j=\widetilde{Q}^a_i Q_{a}^j,
\ee
where we have defined the gauge invariant meson $M_i^j$. The effective low energy superpotential will consist of this classical piece as well as non-perturbative effects which will depend on $M_i$, the number of flavors charged under $U(N_i)$ (for a review see \cite{Intriligator}).  Quantum effects will become important at a scale $\Lambda_1$ determined by the matching relation $\Lambda_1^{3N_i-M_i}=\mu^{N_i} \Lambda^{2N_i-M_i}$. To find the vacua of the $\CN=1$ theory, we  need to extremize this effective superpotential. 
This problem has been studied in various papers \cite{Hori:1997ab,Carlino:2000uk,deBoer:1997ap}. 
Here, we will simply summarize the results. 

\subsubsection*{\underline{$\mathbf{U(1)}$ with $\mathbf{M_i \geq 2} $ flavors}:} We start with a  simple  example. Since the theory is IR free, we can use the tree-level superpotential to determine the vacuum structure. After the adjoint is integrated out, the superpotential is 
\be
W=-{1\over 2\mu} (\widetilde{Q}_i Q^i)^2. 
\ee
There are two different solutions to the F and D-flatness conditions:
\begin{enumerate}
\item  $Q =\tilde{Q}=0$, which leaves $U(1)$ unbroken (we will  call this the $r=0$ vacuum).
\item
$U(1)$ is Higgsed by nonzero VEVs of squarks.
By flavor symmetry, we can set $Q=[K,0,0,...,0]$ and 
$\widetilde{Q}=[0,K,0,...,0]$. The unbroken flavor symmetry is
$SU(M_i-2)\times U(1)$.   We will call  this $r=1$ vacuum since the squark VEVs match the form of the $r=1$ Higgs branch described in the previous section.
\end{enumerate}

\subsubsection*{\underline{$\mathbf{U(N_i) ~{\mathrm \bf{with}}~ M_i< N_i}$ flavors}:}

For $M_i<N_i$, non-perturbative $SU(N_i)$ dynamics generates an Affleck-Dine-Seiberg (ADS) superpotential \cite{ADS}. Hence the effective superpotential is 
\begin{equation}
\label{Quantum-1}
W=-{1\over 2\mu} \Tr(M^2) +(N_i-M_i) \Lambda_1^{ 3N_i-M_i \over N_i-M_i}
(\det(M))^{-{1\over N_i-M_i}},
\ee
where the second term is the ADS superpotential. 
The F-flatness condition yields 
\be
-{1\over \mu} M - \Lambda_1^{ 3N_i-M_i \over N_i-M_i}(\det(M))^{-{1\over N_i-M_i}}
M^{-1}=0,
\ee
or
\be
M^2=-\mu ({\Lambda^{N_i} \over \det (M)})^{1\over N_i-M_i}.
\label{M-eq}
\ee
It was shown in \cite{Hori:1997ab} that although $M$ cannot be brought to a diagonal form by a $U(M_i)$ transformation, it can  be diagonalized in $GL(M_i,C)$, and that (\ref{M-eq}) can be solved by such a diagonal $M$ with at most two different entries:
\bea
M&=&\diag \Bigl(\alpha_1,  \dots \alpha_1, \alpha_2, \alpha_2 \dots \alpha_2 \Bigr). \label{sol-M}\\
&  & ~~~~~~~ \underbrace{  ~~~~  r ~~~~  } ~~~~~~\underbrace{M_i-r}  \nonumber 
\eea
The moduli space of vacua  then consists of $GL(M_i,C)$ orbits through the diagonal solution. It is the space ${GL(M_i, C) \over GL(r, C) \times GL(M_i-r, C)}$. 
Putting  this ansatz into (\ref{M-eq}) shows that $\alpha_2=-\alpha_1$ with
\be \label{alpha-1}
\alpha_1= (-)^{M_i-r\over 2N_i- M_i} \mu \Lambda.
\ee 
Here we can take $r\leq \lfloor M_i/2 \rfloor $ since the case with $r>  \lfloor M_i/2 \rfloor$ is related to  $r\leq  \lfloor M_i/2 \rfloor$  by permutation of the diagonal entries which is an element of $GL(M_i,C)$.
For $r<M_i/2$, there are $2N_i-M_i$ vacua.\footnote{Strictly speaking we should say that there are $2N_i - M_i$ orbits of $GL(M_i,C)$ which are all vacua.}  We will show in Sec.~\ref{sec:strong} that 
this number can also be understood from 
the underlying $\CN=2$ theory.
For $r={M_i  \over 2}$,  (\ref{alpha-1}) shows that the solutions come in pairs that are related by
sign flips.    Each pair is related by  
a permutation of diagonal entries, which is an element of $GL(M_i, C)$. Therefore, there are only $N_i -{M_i \over 2}$ vacua for $r=M_i/2$. 
In each of the vacua, the unbroken gauge symmetry is $U(N_i - M_i)$.  At the special point along the $GL(M_i,C)$ orbits where the squark VEV looks like (\ref{alpha-1}) the unbroken flavour symmetry is ${U(r) \times U(M_i - r) \over U(1)}$ but at generic points in the orbit the unbroken flavor symmetry is  $SU(M_i-2r) \times U(1)^r$. 
The gauge symmetry is broken down to $U(N_i-M_i)$ with
no light fields charged  under $U(N_i - M_i)$.  Therefore, the $SU(N_i-M_i)$ part will confine  leaving only a $U(1)$ factor in the IR.

\subsubsection*{\underline{$\mathbf{U(N_i) ~\mathrm{\bf{with}} ~M_i=N_i}$ flavors}:}

For $M_i=N_i$, the $SU(N_i)$ factor confines and the low energy degrees of freedom are the mesons $M_i^j$ and baryons $B$ and  $\widetilde{B}$. The IR theory is a non-linear $\sigma$-model with $M$ and $B$ satisfying the constraint $\det(M)-B\widetilde{B}=\Lambda_1^{2N_i}$ \cite{Seiberg:1994bz}. In addition, there is a $U(1)$ which survives at low energies and the baryons  of $SU(N_i)$ are charged under it.  The constraint between $M$ and $B$ can be implemented by introducing a Lagrange multiplier $X$. The effective superpotential is:
\be \label{Quantum-2}
W=-{1\over 2\mu} \Tr(M^2)+X (\det(M)-B\widetilde{B}-\Lambda_1^{2N_i}).
\ee
The $\CN=1$ vacua can be found by solving the F-term  conditions derived from this superpotential. 
Following \cite{Hori:1997ab}, when $B = \tilde{B} = 0$, the solutions are given in terms of a diagonal meson matrix $M$ as in  (\ref{sol-M}) with 
\be
\alpha_1=(-)^{M_i-r \over M_i} \mu \Lambda, 
\ee
and $\alpha_1 = -\alpha_2$.  Thus, for $r < M_i/2$ there are $2N_i- M_i = M_i$ vacua while for $r = M_i/2$ there are  $N_i- M_i/2$  vacua, following the reasoning given in the $M_i < N_i$ discussion above.\footnote{Again, we should really be talking about orbits of $GL(M_i,C)$ rather than discrete vacua.} 
The overall $U(1)$ is not higgsed and therefore survives in the IR.
There is another vacuum where $M=X=0$,  $|B|=|\widetilde{B}|$ and $B \widetilde{B}=-\Lambda^{2N_i}$.   The overall $U(1)$ is Higgsed in this vacuum, which will turn out to be the weak coupling limit of a branch appearing in the strong coupling analysis that we will later call the ``baryonic branch".

\subsubsection*{\underline{ $\mathbf{U(N_i) ~{\mathrm{\bf{with}}}~ M_i\geq N_i+1}$  flavors }:}
For this case, the appropriate low energy degrees of freedom are the dual quarks and mesons of the magnetic theory with gauge group $U(\widetilde{N}_i)$ and $\widetilde{N}_i = M_i - N_i$.   The superpotential is (details can be found in \cite{Hori:1997ab,Carlino:2000uk})
\be
W=-{1\over 2\mu} \Tr(M^2)+{1\over \lambda} \widetilde{q} M q +(N_i-M_i)({\Lambda_1^{3N_i-M_i} 
\over \det(M)})^{1\over N_i-M_i}.
\ee
Let us consider $M_i < 2 N_i$.   It turns out that there are then two different cases that we need to discuss separately.  If the meson matrix is non-degenerate (i.e., has rank $M_i$), it is straightforward to show that the meson $M$ is given as  in (\ref{sol-M}) with $\alpha_1 = -\alpha_2$ as in (\ref{alpha-1}).  Thus, there are $2N_i-M_i$ such vacua.  Effectively, the meson field $M$ gives masses to all dual quarks so that at low energies we get a  $U(\widetilde{N}_i)$ gauge theory. The $SU(\widetilde{N}_i)$ factor confines, leaving a decoupled $U(1)$ in the IR.   On the other hand, 
if the meson matrix $M$ is degenerate (for details, see  \cite{Carlino:2000uk}), there are various vacua labeled by an index $l =0,1,..,\widetilde{N}_i$.\footnote{In  \cite{Carlino:2000uk}, $l$ runs from $0$ to $\widetilde{N}_i - 1 $ since they are working with $SU(N)$ rather than $U(N)$ theories.  Also as emphasized  in \cite{Carlino:2000uk}, the counting of vacua that we are using is appropriate when we approach the massless limit for massive flavors.  This is indeed  what we are doing in this paper.}
(In effect, these are the $r$'th Higgs branches, which we defined in Sec. \ref{classical}, appearing now for the dual quarks.)   In the $l$'th branch, the dual quarks Higgs 
the gauge group down: $U(\widetilde{N}_i) \rightarrow U(\widetilde{N}_i-l)$.   The meson $M$ gives masses to the remaining quarks, leaving no light matter fields charged under $U(\widetilde{N}_i-l)$. As a result, the $SU(\widetilde{N}_i-l)$ factor will confine leaving $U(1)$ in the IR.  
Thus, for each $r < \widetilde{N}_i$  there are   $\widetilde{N}_i-l$ vacua.      There is one more case, with $r = \widetilde{N}_i$, for which there is a single vacuum with no surviving gauge group at low energies.  As we will see, this vacuum arises as the weak coupling limit of a ``baryonic branch'' in the strongly coupled region.

\subsubsection{\textbf{Counting of vacua}}
Using the above analysis, we can count the total number of vacua in a theory in the various $r$'th branches. Let us summarize the results.  For $M_i\leq N_i$, there are $(2N_i-M_i)$ vacua  for each $r$'th branch. For $M_i\geq N_i+1$,  if $r \geq M_i-N_i$ there are $(2N_i-M_i)$ vacua. However,  when
$r<M_i-N_i$, there are $(M_i-N_i-r)$ additional vacua, leading to a total of $(2N_i-M_i)+(M_i-N_i-r)=N_i-r$ vacua. 
In addition, we need to discuss three special cases separately. The first two  are
baryonic branches at $r=N_i$ or $r=M_i-N_i$ and the third  is the $r=N_i-1$ non-baryonic
branch. All of these three cases have one vacuum.

In summary,
\be
\label{counting}
\#~~of~~vacua=\left\{  \begin{array}{ll} 
1, & ~~~~~r=N_i-1, \\
1, &  ~~~~~r=M_i-N_i, N_i~~(\mathrm{baryonic}),  \\
2N_i-M_i,& ~~~~~r\geq M_i-N_i, \\
N_i-r,&  ~~~~~r< M_i-N_i. \\
\end{array}
\right.
\ee

In this subsection,  we have enumerated the quantum mechanical vacuum structure  of $\CN = 1$ $U(N_i)$ gauge theories with some massless flavors.   A variety of different phenomena including confinement, Higgsing, the  generation of non-perturbative superpotentials etc. contributed to the details of the analysis.   Below we will see that the vacua that we have enumerated, their symmetries and general properties match the structure of various vacua in the strongly coupled theory.   This will raise the possibility of interpolating smoothly between the weakly and strongly coupled regions.

\subsection{\textbf{Strong coupling analysis}}
\label{sec:strong}
In Section \ref{classical}, we discussed the various $r$'th Higgs branches and their roots, which are the points where the $r$'th Higgs branch meets the Coulomb branch. Our discussion mostly focused on a single factor $U(N_i)$ of the gauge group $\prod_{i=1}^k U(N_i)$. In the following, we will continue to mostly focus on one of the gauge group factors. We need to find  points on the $\CN=2$ moduli space which are not lifted by the $\CN=1$ deformation. As we will see in detail below, these are special points on the roots of the $r$'th Higgs branches where a certain number of magnetic monopoles become massless. We summarize these points in Table~\ref{strong-table}, including the unbroken low energy gauge group at the root of the Higgs branch in the $\CN=2$ theory, the number of massless monopoles which will acquire VEVs because of the $\CN=1$ deformation, the number of unbroken $U(1)$'s left after the $\CN=1$ deformation, and the form of the SW curve at these special points. The table will be explained in greater detail in the subsections below. 
{\scriptsize
\TABLE[h]
{\begin{tabular}{|c|c|c|c|} \hline
  & Generic $r$ &  $r=N_f-N
$  & $r=N$ \\ \hline
${\cal N}=2$ gauge group  & &&\\
at the root  & $SU(r)\times U(1)^{N-r}\times U(1)_B$  & 
 $SU(r)\times U(1)^{N-r}\times U(1)_B$  & $SU(N)\times U(1)_0$   \\
\hline
Condensed monopoles &  $N-r-1$  &  $N-r$  &  $0$ \\ \hline  
Unbroken $U(1)$ after &&& \\ $\CN=1$ deformation   &  $1$  &  $0$ &  $0$   \\ \hline 
SW curve  & &&\\
$y^2=$ & $~(x+m)^{2r}H_{N-r-1}^2(x)F_{2}(x)$  & 
$~(x+m)^{2r}H_{N-r}^2(x)$ & $~f(q)(x+m)^{2N}$ \\ \hline 
\end{tabular}
\caption{Structure of the roots of Higgs branches in a $U(N)$ theory with $N_f$ flavors.   In all cases $r \leq N$ and $r\leq N_f/2$.   The generic $r$'th branch in the first column exists for all $N_f$ and $N$.   At special points on the roots of these branches at most $N - r - 1$ monopoles can become  massless.   When $N_f \geq N$ the additional branches indicated in the second and third columns are also possible.   At special points on the roots of these ``baryonic branches'' an extra monopole will become massless leading to a total of  $N - r$ massless monopoles.  
We will refer to branches with $N - r -1$ condensed monopoles as ``non-baryonic'' branches.
\label{strong-table}}}
}
\subsubsection{\textbf{Which points are not lifted by the $\CN=1$ deformation}}

Consider a $U(N_i)$ theory with $M_i$ flavors.  We will first discuss the generic $r$'th branch in the first column of Table~\ref{strong-table} and then proceed to the special cases in the other two columns.  Classically, at a generic point on the $r$'th Higgs branch, the gauge symmetry is $U(N_i-r)$, while at the root, it is enhanced to $U(N_i)$. Quantum mechanically, in the $\CN=2$ theory (with $W(\Phi)=0$), the enhanced gauge symmetry at the root of the $r$'th Higgs branch is $SU(r) \times U(1)^{N_i-r}\times U(1)_B$.  There are massless flavors charged under the group $SU(r)$ and one of the $U(1)$'s. In addition, at special points on the root, there are $N_i-r-1$ monopoles which become massless. At these points, the theory has a spectrum\cite{Argyres:1996eh}:
\be \label{effect-r-th}
\begin{array}{cccccccccccc}
 &  SU(r) & \times  &  U(1)_0 & \times  &  U(1)_1 & \times & \cdots & \times
& U(1)_{N_i-r-1} & \times & U(1)_B \\
 M_i\times q &  {\bf r} &   &  1 &   &  0 &  & \cdots & 
& 0 & & 0  \\
 e_1 &  {\bf 1} &   & 0 &   &  1 &  & \cdots & 
& 0 & & 0  \\
 \vdots &  \vdots &   & \vdots &   &  \vdots &  & \ddots & 
& \vdots & & \vdots \\
 e_{N_i-r-1} &  {\bf 1} &   & 0 &   &  0 &  & \cdots & 
& 1 & & 0  \\
\end{array}
\ee
Here $q$ are the massless quarks and the $e_i$ are the massless monopoles and we have indicated the charges carried by these fields under the various product gauge group factors.

We will now see that the points which are not lifted by an $\CN=1$ deformation are precisely those  in the root of the $r$'th Higgs branch, where $N_i-r-1$  monopoles become massless.    (There are $2N_i - M_i$ such points with massless monopoles which are related to each other by the $Z_{2N_i - M_i}$ R-symmetry of the $\CN = 2$ theory.   This can be seen by examining the $\CN = 2$ Seiberg-Witten curve \cite{Argyres:1996eh}.)   We could consider general superpotentials of order $k$: $W(\Phi) = \sum_{i=1}^k {g_i \over k} \Tr(\Phi^k)$.  However, here we are really interested in analyzing the dynamics in one of the gauge group factors in a $U(N)$ theory that is broken to a product of $U(N_i)$ factors in the weak coupling region.  We will see later that for this goal it will always suffice to just consider a quadratic superpotential of the form $W(\Phi)= \mu \Tr \Phi^2$.
Then, around points where $N_i - r - 1$ monopoles are becoming light, the low energy effective action is \cite{Argyres:1996eh}:
\bea \label{non-b-1}
W & = & \sqrt{2} \Tr(q \phi \widetilde{q})+\sqrt{2} \psi_0 \Tr(q\widetilde{q})
+\sqrt{2}\sum_{k=1}^{N_i-r-1} \psi_k e_k\widetilde{e}_k\\
& &  +\mu
(\Lambda \sum_{i=0}^{N_i-r-1} x_i\psi_i+{1\over 2} \Tr \phi^2+
{1\over 2}\psi_B^2), \nonumber
\eea
where $\phi$ is the adjoint of $SU(r)$ and $\psi_k$ is the adjoint of 
other $U(1)$ factors. The terms on the second line arise from the mass deformation $\mu \Tr(\Phi^2)$
written in terms of the low energy fields with constants $x_i\sim 1$. 
There is no light field charged under $U(1)_B$, which will remain decoupled in the IR. Notice that
$\psi_k,~k=1,....,N_i-r-1$ and $U(1)_B$ do not interact with the  $SU(r)\times U(1)_0$ factor. Thus the 
conditions for F-flatness can be separately solved for these two parts.  
For  $e_k,\widetilde{e}_k, \psi_k,~k=1,..,N_i-r-1$ and $\psi_B$, the F-term
equations are
\be
0=\psi_k e_k= \psi_k\widetilde{e}_k=\sqrt{2}e_k\widetilde{e}_k+\mu 
\Lambda x_k=\psi_B.
\ee
With  $\mu\neq 0$, the solution has
$\psi_k=0$ and $|e_k|=|\widetilde{e}_k|$ with $
 \sqrt{2}e_k\widetilde{e}_k=-\mu \Lambda x_k$. This implies that  $U(1)^{N_i-r-1}$ is completely Higgsed down in the  IR. The $U(1)_B$ is left unbroken. 

For the $SU(r)\times U(1)_0$  factor, the D-term equations are:
\bea
[\phi,\phi^\dagger]=0= q^i_a (q^\dagger)^b_i-(\widetilde{q}^\dagger)^i_a
\widetilde{q}^b_i,
\label{3.4}
\eea
and the F-terms yield
\bea
0 & = & (\phi^a_b+\psi_0 \delta^a_b) q^i_a=\widetilde{q}^b_i (
\phi^a_b+\psi_0 \delta^a_b), \\
0 & = & \sqrt{2} q^i_a \widetilde{q}^a_i+\mu \Lambda x_0, \\
0 & = &  \sqrt{2} q^i_a \widetilde{q}^b_i-{\sqrt{2}\over r} \delta_a^b 
(q^i_c \widetilde{q}^c_i)+\mu \phi_a^b.
\label{3.7}
\eea
 By a redefinition $\widetilde{\phi}^a_b=\phi^a_b+\psi_0 \delta^a_b$, we can reduce these equations 
to the analysis in \cite{Argyres:1996eh}.\footnote{We can simply set $\nu=0,\rho\neq 0$ and $r\leq 2M_i$ case in equations (2.9), (2.16) and (6.5) of \cite{Argyres:1996eh}.}  This gives:
\bea
\psi_0 & = & {\Lambda x_0 (l-r) \over lr},~~~~~~~~~~~~~~~~~~\phi=\left[ \begin{array}{cc}
-\psi_0 I_{l} & 0 \\ 0 & -{\Lambda x_0 \over r} I_{r-l} \end{array} \right], \label{small-q} 
\\
q & = & \left[ \begin{array}{ccc} K_{l\times l} & 0_{l\times l} & 0_{l\times (M_i-2l)} \\
 0_{(r-l)\times l} & 0_{(r-l)\times l} & 0_{(r-l)\times (M_i-2l)} \end{array} \right],~~~~~~
~^t\widetilde{q}  =  \left[ \begin{array}{ccc} 
\widetilde{K}_{l\times l} & \lambda_{l\times l} & 0_{l\times (M_i-2l)} \\
 0_{(r-l)\times l} & 0_{(r-l)\times l} & 0_{(r-l)\times (M_i-2l)} \end{array} \right].
\nonumber 
\eea
If $l < r$, the gauge group is Higgsed down to $SU(r-l) \times U(1)$, with no charged light fields. The $SU(r-l)$ factor then confines, leaving the $U(1)$ factor at low energies. This, and the decoupled $U(1)_B$ mentioned above, give two $U(1)$s at low energies. For the $l=r$ case, the $SU(r) \times U(1)_0$ group is completely Higgsed leading to just a single $U(1)_B$ in the IR.
 In fact, it is argued in \cite{Carlino:2000uk} 
that the solutions with $l<r$, 
lie
beyond the validity of the low energy effective action since they suffer
fluctuations $\psi_0\sim \Lambda$ much larger than $\mu$.
So  they  cannot be trusted and  only
the $l=r$ case should be considered.\footnote{The authors of \cite{Carlino:2000uk}  argue that the $l<r$ vacua should not exist on the basis of comparing the number of different vacua expected in the weak coupling region as compared to the strong coupling discussion here.}  The unbroken global flavor symmetry is 
$ SU(M_i-2r)\times U(1)^r$.

\paragraph{The case of ${\mathbf{r=M_i-N_i}}$:}  When the number of flavors $M_i$ exceeds the number of colours $N_i$, an $SU(N_i)$ admits so-called baryonic Higgs branches where the baryons have VEVs.   In a $U(N)$ theory, there are no gauge-invariant baryons.  Nevertheless the analogue of the baryonic $SU(N)$ branches exists in this case also.   
Consider and $r$'th Higgs branch with $r = M_i - N_i$.   At the root of this branch there are points where $N_i - r -1$ monopoles become massless.  However, there is an additional point at the root of this Higgs branch, where $N_i -r$ monopoles become massless.   (This can be shown by analysis of the Seiberg-Witten curve in this case, as we will explain further below, and have summarized in Table~\ref{strong-table}.)
This additional point is invariant under the $Z_{2N_i-M_i}$ R-symmetry group and the surviving gauge symmetry is 
 $U(r)\times U(1)^{N_i-r}$ with $r=M_i-N_i$. The  spectrum of massless particles is:
 \be \label{effect-bary-th}
\begin{array}{cccccccccccc}
 &  SU(r) & \times  &  U(1)_1 & \times  &  U(1)_2 & \times & \cdots & \times
& U(1)_{N_i-r} & \times & U(1)_B \\
 M_i\times q &  {\bf r} &   & 1/r &   &  1/r &  & \cdots & 
& 1/r & & -N_i/r  \\
 e_1 &  {\bf 1} &   & -1 &   &  0 &  & \cdots & 
& 0 & & 0  \\
 \vdots &  \vdots &   & \vdots &   &  \vdots &  & \ddots & 
& \vdots & & \vdots \\
e_{N_i-r} &  {\bf 1} &   & 0 &   &  0 &  & \cdots & 
& -1 & & 0  \\
\end{array}
\ee
where $q_i$ are the quarks and $e_i$ are the magnetic monopoles.
The low energy effective superpotential  for these particles is governed by $\CN=2$ supersymmetry, which is broken by addition of the mass term $\mu \Tr \Phi^2$:
\bea \label{non-b-2}
W & = & \sqrt{2} \Tr(q \phi \widetilde{q})+{\sqrt{2}\over r} \Tr(q\widetilde{q})
(\sum_{k=1}^{N_i-r} \psi_k)-{\sqrt{2} N_i\over r} tr(q\widetilde{q})\psi_B\\
& & -\sqrt{2}\sum_{k=1}^{N_i-r} \psi_k e_k\widetilde{e}_k +\mu
(\Lambda \sum_{i=1}^{N_i-r} x_i\psi_i+{1\over 2} tr \phi^2
+{1\over 2} \psi_B^2).\nonumber
\eea
In this case, the gauge field $U(1)_B$ is not decoupled. Furthermore, $\psi_k$ is not decoupled from  $q,\widetilde{q}$. The F-term of $\psi,e,\widetilde{e}$ and $\psi_B$  yields
\be
0=\psi_k e_k= \psi_k\widetilde{e}_k={\sqrt{2}\over r}\Tr(q \widetilde{q})
-\sqrt{2}e_k\widetilde{e}_k+\mu \Lambda x_k=-{\sqrt{2} N_i\over r}\Tr(q \widetilde{q})
+\mu \psi_B.
\ee
If $\mu\neq 0$, we need to consider two cases separately: 
\begin{enumerate}
\item  ${\psi_B\over N_i}+\Lambda x_k \neq 0,~~\forall k$ . This implies that
\be \psi_k=0, ~~~~ e_k,\widetilde{e}_k \neq 0 ~~~{\mathrm for} ~~~~ k=1,...,N_i-r ~~~
\mathrm{and}  ~~~~~\Tr(q\widetilde{q})={ \mu \psi_B r \over \sqrt{2} N_i}.\ee
{\em i.e.}  $U(1)^{N_i-r}$ is higgsed. The remaining
equations for $q,\widetilde{q}, \phi$ are:
\bea
0 & = & [\phi,\phi^\dagger]=q^i_a (q^\dagger)^b_i
-(\widetilde{q}^\dagger)^i_a\widetilde{q}^b_i, \\
0 & = & \sqrt{2} q^i_a \widetilde{q}^b_i-{\sqrt{2}\over r} \delta_a^b 
(q^i_c \widetilde{q}^c_i)+\mu \phi_a^b,\\
0 & = & (\phi^a_b-{N_i\over r}\psi_B \delta^a_b) q^i_a=\widetilde{q}^b_i (
\phi^a_b-{N_i\over r}\psi_B \delta^a_b), \\
0 & = & \sqrt{2}\Tr(q\widetilde{q})-{ \mu \psi_B r \over  N_i}.
\eea
These equations are exactly the same equations for general $r$-th
branch (\ref{3.4}, \ref{3.7}) with $-N_i \psi_B \rightarrow \psi_0$ and $-{\psi_B r\over N_i}
\rightarrow x_0$. Hence the general solution is given by (\ref{small-q}).   In the general $r$-th branch, the authors of  \cite{Carlino:2000uk}  argued that the solutions in (\ref{small-q}) with $l<r$ cannot be trusted and do not exist quantum mechanically.  In the present case, however, all the solutions in (\ref{small-q}) with both $l<r$ and $l=r$ are valid
\cite{Carlino:2000uk} because all $\psi_k,\psi_B\sim  0$ and therefore are far from the dynamical scale of the theory.   Following the discussion below (\ref{small-q}), the $l<r$ vacua have a $SU(r-l) \times U(1)$ at low energies and the $SU$ factor confines leaving $r-l$ discrete vacua.   These vacua are the strong coupling limit of the situations with degenerate meson fields $M$ in the weak coupling analysis  in Sec.~\ref{weak} of the cases with more flavors than colors where we found $\widetilde{N}_i - l = M_i - N_i - l$ vacua.   For $l<r$ although the $U(1)^{N_i-r}$ facotr is higgsed,
there is one decoupled $U(1)\subset U(r)$ left while for $l=r$, there is no
$U(1)$ left and the whole theory is higgsed.
\item ${\psi_B\over N_i}+\Lambda x_{k_0} = 0$ for a particular
$k_0$ (because in general all $x_k$ are different). 
\newline Then,  we can have 
$e_{k_0}=\widetilde{e}_{k_0}=0$ with nonzero $\psi_{k_0}$. Thus, one of the monopoles does not condense in this case and 
only $N_i-r-1$ of the $U(1)$ factors  will be Higgsed down.  The remaining
equations for $q,\widetilde{q}, \phi$ in this case will be similar to
the generic $r$'th branch with the role of $\psi_B$ replaced by $\psi_{k_0}$.  Since only $N_i - r -1$ monopoles are condensed this is in fact one of the generic $r$'th branches that we discussed earlier in this section.   There are $N_i-r$ ways to choose $k_0$ which leads to $N_i-r=2N_i-M_i$. This matches the weak coupling counting. So following the discussion in \cite{Carlino:2000uk}  we once again only keep the $l=r$ solutions in (\ref{small-q}) .
\end{enumerate}
\paragraph{The case of $\mathbf{r=N_i}$:}
This case  also needs a separate  discussion. This is relevant when $M_i \geq 2N_i$. In such a situation, there is an $r=N_i$ branch, whose root is not lifted by the $\CN=1$ deformation.   Since $N_i - r = 0$, no monopoles need to condense to give an $\CN = 1$ vacuum.  That the root of this branch (without any monopole condensation) is not lifted can be shown along the same lines as the above $r=M_i-N_i$ discussion. At a generic point on this branch, the gauge group is completely broken. At the root of this branch, the gauge group is also completely Higgsed down after the $\CN=1$ deformation.  So on this branch, as in the ``baryonic'' $r=M_i-N_i$ case, there is no surviving gauge group at low energies. This is in contrast to the generic $r$'th branches, ($r \neq M_i-N_i$ and $r\neq N_i$) where a $U(1)$ gauge group survived at low energies after the $\CN=1$ deformation.

\subsubsection{\textbf{SW curve factorization and roots of the $\mathbf{r}$'th Higgs branches}}
\label{sec-SW}
As discussed above, the subspace of the $\CN=2$ moduli space not lifted by the $\CN=1$ deformation consists of special points at the roots of the $r$'th Higgs branches where a certain number of mutually local magnetic monopoles are becoming massless. We determine these points by treating the superpotential as a small perturbation around the $\CN=2$ theory and using the Seiberg-Witten curve for an $\CN=2$  $U(N)$ gauge theory with $M_i$ hypermultiplets, with mass $m_i$ \cite
{Klemm, Argyres, Hanany}.
 \be \label{SU-SW}
y^2= P_{N_i}(x)^2-4 \Lambda^{2N_i-M_i}  (x+m_i)^{M_i}=
\prod_{a=1}^{N_i} (x-\phi_a)^2- 4 \Lambda^{2N_i-M_i} (x+m_i)^{M_i}
,~~M_i\leq 2N_i-2.
\ee
Therefore, at the root of the $r$'th Higgs branch, we can write the adjoint scalar as:
\be \label{r-th-Phi}
\Phi=(-m_i,-m_i,-m_i,...,\phi_{1}, ...,\phi_{N_i-r})~.
\ee
Then, for such points,  curve (\ref{SU-SW}) factorizes as
\be  \label{factorize-1}
y^2=(x+m_i)^{2r}\Bigl(P_{N_i-r}^2(x)-4 \Lambda^{2N_i-M_i} (x+m_i)^{M_i-2r}\Bigr)
\ee 
Furthermore, we need to find the points at the roots of these Higgs branches where  $N_i-r-1$ monopoles are massless as in (\ref{effect-r-th}).   For this, we need the second factor in the above curve to factorize as
\be \label{factorize-2}
P_{N_i-r}^2(x)-4 \Lambda^{2N_i-M_i} (x+m_i)^{M_i-2r}=H_{N_i-r-1}(x)^2 F_{2}(x).
\ee
Above we discussed additional ``baryonic'' branches in which $N_i - r$ monopoles can become massless.  At such points the Seiberg-Witten curve factorizes as:
\be \label{factorize-bary}
P_{N_i-r}^2(x)-4 \Lambda^{2N_i-M_i} (x+m_i)^{M_i-2r}=H_{N_i-r}(x)^2.
\ee

\paragraph{Multiple factors and the underlying $U(N)$ picture}
So far, in this section, we have considered one of the factors of the group $U(N_i)$ with a quadratic superpotential for the adjoint field $\Phi_i$ charged under that group. We are now ready to discuss the general picture. 
Classically $U(N)$ with $N_f$ flavors is broken
to $\prod_{u=1}^k U(N_u)$ with $M_u$ (which could be zero) massless flavors charged under each 
factor. This breaking occurs via the Higgs mechanism when the adjoint $\Phi$ acquires a diagonal vev with diagonal entries being roots $a_i$ of the $k$-th order polynomial $W'(x)$. This picture only makes sense  quantum mechanically if the difference between the various $a_i$ is much bigger than $\Lambda$, the dynamical scale of the underlying $\CN=2$ $U(N)$ theory. Let us assume that we are at such a point. Then, the SW curve for the $U(N)$ theory should approximately factorize as
\be
y^2=P_N^2(x)-4\Lambda^{2N-N_f}\prod_{i=1}^k (x+m_i)^{M_i}\sim\prod_{i=1}^k\Bigl(P_{N_i}^2(x)-4\Lambda_i^{2{N_i}-{M_i}} (x+m_i)^{M_i}\Bigr)
\ee
where $\Lambda_i$ are the dynamical scales of each factor $U(N_i)$.
The analysis  above applies to each of these factors. For each factor, the quadratic term in the superpotential will be the only relevant term since higher order pieces will be suppressed by a large scale. After the ${\cal N}=1$ deformation, the unlifted points  on the $\CN =2$ moduli space are special locations on the $r_i$-th Higgs branch root  for each $U(N_i)$ factor such that a  total of $N-\sum_i r_i -k$ monopoles become massless.   (More monopoles could become massless, but in that case, either some of the Higgs branches in question are ``baryonic" or the classical limit has fewer $U(N_i)$ factors.) We can now vary the parameters in the superpotential holomorphically, and move to the region of strong coupling where the description in terms of product gauge groups does not make any sense. However, we will still require the curve to have overall factor $\prod_{j=1}^k(x+m_j)^{2r_j}$ as well as have $N-\sum_i r_i -k$ double roots; {\em i.e.}
\bea
y^2 &= & \prod_{j=1}^k (x+m_j)^{2r_j} \Bigl(P_{N-\sum_{j=1}^k r_j} (x)^2-
4\Lambda^{2N-M} \prod_{j=1}^k (x+m_j)^{M_j-2r_j}\Bigr)\\
&= & \prod_{j=1}^k (x+m_j)^{2r_j} F_{2k}(x)H_{N-k-\sum_{u=1}^k r_u} (x)^2
\label{general-double}
\eea 
This is the generic picture in the $r$'th Higgs branch. There are special cases mentioned above, for example $r_i=M_i-N_i$ where there is a branch on which $N_i-r_i$ monopoles condense, after the $\CN=1$ deformation. At the $\CN=2$ level of the SW-curve, these cases will require an extra double root. A similar story holds for the $r_i=N_i$ branch for cases where $M_i  \geq 2N_i$. There exists a branch, for this special case also, in which $N_i -r_i$ monopoles become massless, instead of the genric number $N_i-r_i-1$. As we will see in the examples below, these different branches will play an important role in understanding the interpolations between vacua of the $\CN=1$ theories obtained in the various classical limits. 

\subsubsection{{$\mathbf{W(\Phi) ~\mathrm{\bf{and}}~ \CN=1}$ \textbf{vacua}}}
Thus far in our discussion of factorization of the SW curve, the superpotential $W(\Phi)$ has not played any role. To find the $\CN=1$ vacua, we need to extremize the superpotential on the subspace of the moduli space which leads to the  factorized form of the SW curve described above (\ref{general-double}). This extremization procedure will yield isolated vacua. At this juncture, we can ask two questions: (i) Given a superpotential, which points on the $\CN=2$ moduli space are $\CN=1$ vacua; (ii) Given a point in the $\CN=2$ moduli space, what superpotential will have that point as its extremum. As we will see later, both these questions will play an important role in later analysis.

\paragraph{Extremization}
For the extremization of the superpotential, the condition (\ref{general-double}) can be incorporated via Lagrange multipliers. Using this method, it was shown in  \cite{Cachazo:2001jy} (for the case without flavors) that the extremization problem has a very simple solution in terms of the factorized form of the Seiberg-Witten curve, 
\be
y^2=P_{N}(x)^2- 4\Lambda^{2N} = H_{N-n}(x)^2 F_{2n}(x),
\ee
where $n$ is the number of $U(N_i)$ factors and $N$ is the overall gauge group rank.   The degree $k$ of $W'(x) $ must be greater than or equal to $n$, and we always look at the case $k =n$.   (The generalization to $n < k$ is straightforward \cite{Cachazo:2001jy,Cachazo:2002zk}.)  For $n=k$  \cite{Cachazo:2001jy}  find
\be
g_{n+1}^2 F_{2n}(x)= W'(x)^2+f_{n-1}(x),
\ee
{\em i.e.} the first $n+1$ coefficients of $F_{2n}(x)$ are determined by $W'(x)$. 
The generalization of this to the theory with flavors is discussed in \cite{Ookouchi:2002be}. 
For completeness, we review the procedure and point out a few subtleties.   

The Seiberg-Witten curve with flavors factorizes as follows at points in the moduli space where there are $l$ mutually local massless monopoles:
\be
y^2=P_{N}(x)^2 - 4 \Lambda^{2N-N_f} A(x)=H_l(x)^2 F_{2N-2l}(x),~~~~~
A(x)=\prod_{j=1}^{N_f} (x+m_j),
\ee
where $H_l(x)=\prod_{i=1}^l (x-p_i)$, where $p_i$ are double roots of the polynomial $P_{N}(x)^2 - 4 \Lambda^{2N-N_f} A(x)$. To extremize the superpotential on the subspace with $l$ massless monopoles, we use the superpotential
\footnote{When writing these equations we have assumed that  that $P_N(x)\pm 2\Lambda^{N-{N_f\over 2}}\sqrt{A(p_i)}$ do not have common factors. This is actually only true for the $r=0$ branch.   However, as we have discussed, for a general $r$'th branch the SW curve factorizes as $y^2 = (x +m)^{2r} \tilde{y}^{2}$.  Then the extremization needs to be done for the factor $\tilde{y}^2$ and the resulting algebra is the same as for the $r=0$ case.   We have also specialized to the case of an  even
number of  flavors, but the generalization to an odd number of flavors is straightforward and gives similar results.}
\be
W_{low}=\sum_{p=1}^{k+1} g_p u_p+\sum_{i=1}^l \Bigg\{L_i (P(p_i)-2\epsilon_i 
\Lambda^{N-{N_f\over 2}}\sqrt{A(p_i)})+Q_i {\partial (P(p_i)-2\epsilon_i 
 \Lambda^{N-{N_f\over 2}}\sqrt{A(p_i)}) \over \partial p_i}\Bigg\}
\ee
where $L_i$ and $Q_i$  are the Lagrange multipliers imposing the constraints which restrict us on the relevant subspace and $p_i$ are the double roots. The variation with respect to $p_i$ yields
\be
Q_i {\partial^2 P(p_i) \over \partial p_i^2 }|_{x=p_i}=0. \ee
For a generic superpotential, we can assume there are no  triple roots (${\partial^2 P(p_i) \over \partial p_i^2 }| _{x=p_i}\neq 0$) which implies $Q_i=0$. Following the arguments in \cite{Cachazo:2001jy}, variation with respect to $u_r$ leads to
\be
g_r=\sum_{i=1}^l \sum_{j=0}^N L_i p_i^{N-j} s_{j-r}.
\ee
Using this expression, we can express $W'(x)$ as (see \cite{Cachazo:2001jy} for more details)
\be
W'=B_{l-1}(x) {P_{N}(x)\over H_l(x)}-x^{-1}
C+{\cal O}(x^{-2}),
\ee
where $C$ is a constant and $B_{l-1}(x)$ is an order $(l-1)$  polynomial. Comparing the leading power of $x$ on the two sides of the above equation, we conclude that  when $l=(N-n)$ 
the function $B_{l-1}$ is in fact  a constant and 
\be \label{flavor-F_2n}
 F_{2n}(x)+{4 \Lambda^{2N-N_f}\prod_{j=1}^{N_f} (x+m_j) \over
\prod_{j=1}^{N-n} (x+p_j)^2}={1 \over g_{n+1}^2}\Bigl( W'(x)^2+ {\cal O}(x^{n-1})\Bigr).
\ee
The second term on the LHS of the above equation will contribute\footnote{This point was overlooked in \cite{Ookouchi:2002be} and will be important in the examples discussed in later section.} to $W'(x)$ when $N_f-2(N-n)\geq n$ or $n\geq 2N-N_f$. \footnote{Geometrically, $F_{2n}(x)$ defines the
curve which describes the quantum behaviour of  the low energy $U(1)^n$.  In a geometric engineering picture of the gauge theories we study, flavors arise from D5-branes wrapping non-compact 2-cycles in a local Calabi-Yau 3-fold.  In addition there are D5-brans wrapped on compact $S^2$s giving rise to gauge fields.  One can blow down the $S^2$ and blow up to an $S^3$ and the D5-branes on the compact cycles are replaced by fluxes.  After this transition the superpotential of the geometrically engineered gauge theory can be computed.  Since the non-compact 2-cycles are left alone in this transition it would apepar that the quantum curve $y^2 = F_{2n}(x)\sim
W'(x)^2+f_{n-1}(x)$ should not modified by the presence of flavors.  But the above careful analysis 
(\ref{flavor-F_2n}) tells us that this is not generally true. We will have
\be
F_{2n}(x)=W'(x)+f_K(x),~~~~~~~K=max(n-1,2n+N_f-2N)
\ee
and especially 
\be
{\partial W_{low} \over \partial \log(\Lambda^{2N_c-N_f})} \neq {-b_{n-1} \over
g_{n+1}}
\ee
where $b_{n-1}$ is the coefficient of $x^{n-1}$ in the polynomial $f_K(x)$.
This seems to  make the geometric proof of  Large $N$ duality
 \cite{Cachazo:2002pr,Ookouchi:2002be,Feng:2002gb}
  subtle  in the presence of flavors. 
We will come back to this point in Sec.\ref{sec:multiplication}.
}

The derivation we have sketched above fails in special cases. For example, there are special branches on which we require an extra massless monopole. In such cases, the degree of the function $F(x)$ is actually $(2n-2)$ instead of $2n$, and we need to generalize the above analysis along the lines of  \cite{Cachazo:2002zk,Ahn:2003cq}.

\subsubsection{\textbf{Different branches  with (perhaps) multiple classical limits}}
\label{gen}
Henceforth, for concreteness, we will focus on the case where, in the classical limit, the gauge group $U(N)$ breaks into two factors $U(N_1) \times U(N_2)$.  To study this we will consider a cubic superpotential. For simplicity, we choose all $N_f$ flavors to have the same mass $m_i=m$.  In addition we will choose $N_f = 2l$ to be even for convenience.

At the root of the $r$-th Higgs branch of the $\CN=2$ theory, the SW curve factorizes (as discussed in (\ref{sec-SW})) as
\be
y^2=(x+m)^{2r}(P_{N-r}(x)^2-4\Lambda^{2N-N_f}(x+m)^{N_f-2r}).
\ee
Furthermore, the subspace not lifted by the $\CN=1$ deformation are the points where a total of $N-r-2$ monopoles are becoming massless.  (There are also some ``baryonic'' branches where additional monopoles condense and the analysis needs to be suitably modified in those cases.)
Thus we require $N-r-2$ double roots of the polynomial
\be
P_{N-r}(x)^2-4\Lambda^{2N-2l}(x+m)^{2l-2r}=H_+(x) H_-(x)=H_{N-r-2}(x)^2 F_4(x),
\ee
where $H_{\pm}(x)=P_{N-r}(x)\pm2\Lambda^{N-l}(x+m)^{l-r}$. Suppose $P_{N-r}(-m)\neq 0$. Then, $H_{+}(x)$ and $H_-(x)$ do not have coincident roots. Therefore, we require $s_{+}$ ($s_-$) double roots of $H_+(x)$ ($H_-(x)$), with $s_++s_-=N-r-2$. Thus the different branches are labeled by three integers $(r,s_+,s_-)$. 

On such a factorization locus of the SW curve, the $\CN=1$ superpotential has to be extremized, as explained above. After the extremization, which yields isolated $\CN=1$ vacua, we can take various possible classical limits ($\Lambda \rightarrow 0$, possibly along with other limits of parameters) to transport this $\CN=1$ vacua to the weak coupling region, so that we can interpret these as vacua of a $U(N_1) \times U(N_2)$ theory with flavors charged under the group factors. As we will see later, some branches of the strongly coupled theory will have multiple smooth classical limits.  This will lead to smooth interpolations between vacua of different microscopic theories with the same macroscopic physics, by which we mean that the low-energy gauge group and global symmetries must be the same.  

In the next section we will develop some general relationships between the vacua of different $U(N)$ theories.

\section{Addition and Multiplication maps}
\label{maps}
In \cite{Cachazo:2002zk,Cachazo:2001jy}, a construction (called the {\it multiplication map}) was given which related vacua of a $U(N)$ theory with a given superpotential to vacua of a $U(tN)$ theory with the same superpotential. Hence classical limits of the $U(N)$ theory with gauge group $\prod_i U(N_i)$ are mapped to classical limits of the $U(tN)$ theory with gauge group $\prod_i U(tN_i)$.    This is a powerful observation because in the absence of flavors all confining vacua of higher gauge groups can be analyzed by using the multiplication map to relate to the Coulomb vacua of a lower rank gauge theory.
Below we show that we can find a similar map in the theory with flavors.   What is more, when flavors are present there is an another map, which we will call the {\it addition map}, which relates
certain vacua of a $U(N)$ theory with $N_f$ flavors and $U(N+1)$ theory with $N_f+2$ flavors.   This will also enable us to reduce many analyses to simpler cases.

\subsection{\textbf{Addition map}}
Consider two theories,  $U(N)$ with $N_f$ flavors in the $r$-th branch and 
$U(N')$ with $N_f'$ flavors in  the $r'$-th branch. On these branches the SW curve is
\bea
y_1^2 & = & (x+m)^{2r}\Bigg\{ P_{N-r}(x)^2 -4 \Lambda^{2N-N_f}
(x+m)^{N_f-2r} \Bigg\},
\label{N1-M1-r1} \\
y_2^2 & = & (x+m)^{2r'}\Bigg\{ P_{N'-r'}(x)^2 -4 \Lambda^{2N'-N_f'}
(x+m)^{N_f'-2r'}\Bigg\}.
\label{N2-M2-r2}
\eea
If 
\be \label{two-theory}
N-r=N'-r',~~~~~2N-N_f=2N'-N_f',~~~~~N_f-2r=N_f'-2r',
\ee
the polynomials inside the braces in these two equations are the same. Let $P_{N-r}(x)=P_{N'-r'}(x)=
\det(x-\tilde{\phi})$. Then, on the $r$-th branch of the $U(N)$ theory with $N_f$ flavors, 
the gauge invariant operators $u_k={1 \over k} \Tr \Phi^k$ will be given by 
$u_k={1 \over k}r(-m)^k+\tilde{u}_k$ where $\tilde{u}_k=\Tr \tilde{\phi}^k$. Similar expressions hold for $U(N')$ theory with $N_f'$ flavors in the $r'$ th  branch. Since the $u_k^\prime$s differ from $\tilde{u}_k$s by a shift by a 
constant, the extremization of the superpotential will yield the same solution for the expressions within the braces.  In other words, the solutions for $u_k$ and $\tilde{u}_k$ are the same.   This shows that certain vacua of  $U(N)$ with $N_f$ flavors in the $r$ branch can be related  to those of $U(N')$ with $N_f'$ flavors in the $r'$ branch.

For example, if on the $r$-th branch, the $U(N)$ theory with $N_f$ flavors has a classical limit $U(N_1) \times U(N_2)$  with $N_f$ flavors charged under the $U(N_1)$ factor, then a $U(N+d)$ theory with $N_f+2d$ flavors  will have a classical limit $U(N_1+d) \times U(N_2)$ with $N_f+2d$ flavors charged under $U(N_1+d)$  on the $r+d$ branch. We can also start with a $U(N)$ theory with no flavors. If 
this theory has a classical limit with gauge group $U(N_1) \times U(N_2)$, via the addition map, 
we can conclude that a $U(N+d)$ theory with $2d$ flavors can have a classical limit $U(N_1+d) \times
U(N_2)$ with $2d$ flavors under $U(N_1+d)$. It can also have a classical limit $U(N_1) \times U(N_2+d)$ with $2d$ flavors now charged under $U(N_2+d)$.

The addition map can be used to reduce the study of most (but not all)
$r$-th branches of $U(N)$ with
$N_f$ flavors  to the $r=0$ branch of $U(N-r)$ with $(N_f-2r)$ flavors.   
An example of a vacuum where this cannot be done appears in
a $U(3)$ gauge theory with two flavors which is related by the addition map to a $U(2)$ theory with no flavors.    The $U(3)$ has a classical limit in which $U(3)$ 
reduces to $U(1) \times U(2)$ with  two flavors charged under
the $U(1)$.  However the $U(2)$ theory with no flavors only has a $U(1) \times U(1)$ vacuum.  Applying the addition map to this will lead to a $U(2) \times U(1)$ with two flavors under the $U(2)$.  Thus we cannot recover the $U(3)$ vacuum with flavors charged under $U(1)$ from the addition map.
It is straightforward to show that such special cases can be reduced
via the addition map
to study of an $r=1$ branch of a $U(1)$ factor with some flavors. 

The quantities $(2N-N_f)$ and $N-r$ are invariant under the addition map.
These are exactly the number of vacua in (\ref{counting}). This shows that the counting of vacua in Sec.~\ref{weak} is consistent with the addition map.\footnote{We expect this for the following reason.  If we considered a quadratic superpotential, these quantities count the number of vacua we expect in the weak coupling region (\ref{counting}), and hence also the total number of $\CN = 1$ vacua obtained in the strong coupling region.   But the number of strong coupling vacua should be invariant under the addition map.}

\subsection{\textbf{Multiplication map}}
\label{sec:multiplication}
We now discuss how to relate vacua of a $U(N)$ theory on the $r=0$ branch to vacua
of a $U(tN)$ theory also on the $r=0$ branch. Our discussion is a simple generalization
of the argument in \cite{Cachazo:2001jy} and also be extended to $r>0$ branches by using the addition map.

The Seiberg-Witten curve for a $U(N)$ theory with $N_f$ flavors of bare mass $m=0$
(we set $m=0$ for simplicity) on the $r=0$ branch is:
\be
y^2=P_N(x)^2-4 \Lambda^{2N-N_f} x^{N_f}.
\ee
We assume that $P_N(0) \neq 0$. Also, we will restrict ourselves to an even number of flavors, {\em i.e.} $N_f=2l$ ($l \leq N$ for the $\CN=2$ theory not to be IR free). 
If we are restricted to the subspace of the moduli space where $N-n$ magnetic monopoles become
massless, the curve factorizes as follows:
\be \label{assuming}
P_N(x)^2-4 \Lambda_0^{2(N-l)} x^{2l}=H_{N-n}(x)^2 F_{2n}(x).
\ee
Given such a factorization, we can factorize the SW curve for  a $U(tN)$ with $2tl$ flavors. In this construction, Chebyshev polynomials play a crucial role. These polynomials, of the first and
second kind of degree $t$ and $t-1$ respectively are defined as: 
\be \label{Chebyshev}
{\cal T}_t(x)=\cos(tx),~~~{\cal U}_{t-1}(x)={1\over t}{\partial {\cal T}_t(x)
\over dx},~~~{\cal T}_t(x)^2-1=(x^2-1) {\cal U}_{t-1}(x)^2.
\ee
Our construction will closely follow \cite{Cachazo:2002zk,Ahn:2003cq} (where the case without flavors was discussed), so we will give a rather brief discussion here, emphasizing the new issues which arise when we include flavors.   We start  by defining
\cite{Cachazo:2002zk,Ahn:2003cq}
\be \label{wide-x}
\widetilde{x}={P_N(x)\over 2 \eta \Lambda^{(N-l)} x^l}.
\ee
Then, we can show that if the curve for $U(N)$ factorizes as in (\ref{assuming}), the curve for
$U(tN)$ will factorize as
\be \label{assuming2}
P_{tN}(x)^2-4 \Lambda^{2t(N-l)} x^{2lt}=H'_{tN-n}(x)^2 F_{2n}(x),
\ee
if we take
\be \label{PtN}
P_{tN}(x)= 2(\eta \Lambda^{(N-l)} x^l)^t {\cal T}_t(\widetilde{x}),
\ee
with $\eta^{2t}=1$. To show this, notice that
\bean
 P_{tN}(x)^2-4 \Lambda^{2t(N-l)} x^{2tl}&=&4 \Lambda^{2t(N-l)} x^{2tl}
[{\cal T}_t(\widetilde{x})^2-1], \\
& = & 4 \eta^{2t} \Lambda^{2t(N-l)} x^{2tl}(\widetilde{x}^2-1)
{\cal U}_{t-1}(\widetilde{x})^2,\\ & = & [(\eta \Lambda^{(N-l)} x^{l})^{t-1}
{\cal U}_{t-1}(\widetilde{x})]^2[P_N(x)^2-4 \eta^2\Lambda^{2(N-l)} x^{2l}],\\
& = & [(\eta \Lambda^{(N-l)} x^{l})^{t-1}
{\cal U}_{t-1}(\widetilde{x}) H_{N-n}(x)]^2 F_{2n}(x),
\eean
where we identify $\Lambda_0^{2(N-l)}=\eta^2\Lambda^{2(N-l)}$.
Notice that in the definition of $\widetilde{x}$, there is an $x^l$ in
the denominator. Hence ${\cal U}_{t-1}(\widetilde{x})$ is not really
polynomial in $x$. However, after multiplying it with $x^{l(t-1)}$, we obtain a polynomial. 

By comparing (\ref{assuming}) with (\ref{assuming2}), we see that the factor $F_{2n}(x)$ is the 
same in the two factorizations. Since $F_{2n}$ was related to the superpotential $W'(x)$ after the minimization on the subspace where $N-n$ monopoles are massless, we can relate the $\CN=1$ vacua of the $U(N)$ theory with the $\CN=1$ vacua of the $U(tN)$ theory. 
It was shown in  (\ref{flavor-F_2n}), that the relationship  of  $W'(x)$ to $F_{2n}(x)$ is modified in
certain cases when flavors are added to the theory . Specifically, it was shown that the flavors modify the relationship if $n\geq 2(N-l)$ for a $U(N)$ theory with $2l$ flavors. For a $U(tN)$ theory
with $2tl$ flavors, this condition becomes $n \geq 2t(N-l)$. Since we are considering $l \leq N$, the relationship of $W'(x)$ to $F_{2n}(x)$ maybe  be different for $U(N)$ and $U(tN)$ theories. In fact, when $l <N$ ({\em i.e.} the underlying $\CN=2$ is asymptotically free), we can always take $t$ large enough so that flavors do not modify the relation between $W'(x)$ and $F_{2n}(x)$, {\em i.e.} the relation is universal for $t$ large enough. 
 This is exactly the behavior predicted by geometric transitions and large-N dualities: the non-compact D5 branes should not affect the transition from $S^2$ to $S^3$. 
For the case without flavors, this universal behaviour does not require taking the large $N$ limit. However, when we include flavors, we somehow have to consider a large $N$ limit to recover this universal behavior predicted by geometry\cite{Cachazo:2001jy}.

The recent activity in study of $\CN = 1$ gauge theories has partly arisen from the connection with Matrix models \cite{dv1,dv2,dv3,dglvz}.  WIthin this connection an important quantity is the density function of the quantum mechanical eigenvalues of the adjoint scalar $\Phi$.   This is given by
\cite{Cachazo:2002zk,CDSW}
\be
T(x)={\partial \over  \partial x}\log(P_N(x)+
\sqrt{P_N(x)^2-4 \Lambda^{2(N-l)} x^{2l}})
\ee
and is related to the resolvent $\langle \Tr ({1 \over x-\Phi})\rangle$ in the matrix model computations of the field theory superpotential.
It is easy to show that
\be
T(x)={P'_N(x) \over \sqrt{P_N(x)^2-4 \Lambda^{2(N-l)} x^{2l}}}
-{l\over x} {P_N(x) \over \sqrt{P_N(x)^2-4 \Lambda^{2(N-l)} x^{2l}}}
+{l\over x}
\ee
For the $U(tN)$ theory, we obtain
\bean
P'_{tN}(x) &= & {P'_{tN}(x) \over 2(\eta \Lambda^{(N-l)} x^l)}
-{l\over x}{P_{tN}(x) \over 2(\eta \Lambda^{(N-l)} x^l)}\\
\sqrt{P_N(x)^2-4 \Lambda^{2(N-l)} x^{2l}}& = & 
[(\eta \Lambda^{(N-l)} x^{l})^{t-1}
{\cal U}_{t-1}(\widetilde{x})]\sqrt{P_N(x)^2-4 \Lambda_0^{2(N-l)} x^{2l}}
\eean
which  yields the relation 
\be \label{TTx}
T_t(x)= t T(x)
\ee
In the context of geometric engineering, this simply says that the $U(t N_i)$ is obtained by wrapping $t N_i$ branes  around the $i$-th root of $W'(x)$ while the $U(N_i)$ theory arises from $N_i$ wrapped branes.

As with the addition map, the multiplication map can be used to map vacua of a $U(N)$ theory to vacua of
a $U(tN)$ theory. Since, 
\be
 P_{tN}(x)^2-4 \Lambda^{2t(N-l)} x^{2tl}= [(\eta \Lambda^{(N-l)} x^{l})^{t-1}
{\cal U}_{t-1}(\widetilde{x})]^2[P_N(x)^2-4 \eta^2\Lambda^{2(N-l)} x^{2l}]
\ee
it is easy to see that in the limit $\Lambda\rightarrow 0$, 
$P_{tN}(x)\rightarrow (P_N(x))^t$. $U(N)$ has vacua which have a classical limit in which the unbroken gauge group is $ U(N_1)\times U(N_2)$. These vacua are related to those of a $U(tN)$ theory which have a classical limit $U(tN_1) \times U(tN_2)$.  Also, if 
a vacuum of 
$U(N_1)\times U(N_2)$
with  $2l$ flavors  under $U(N_1)$ can be continuously deformed into a vacuum of  $U(\widetilde{N}_1)
\times U(\widetilde{N}_2)$ with flavors $2l$ under $U(\widetilde{N}_1)$,
then a vacuum of $U(tN_1)\times U(tN_2)$
with flavors $2tl$ under $U(tN_1)$ is continuously deformable to a vacuum of  $U(t\widetilde{N}_1)
\times U(t\widetilde{N}_2)$ with flavors $2tl$ under $U(t\widetilde{N}_1)$.

In \cite{Cachazo:2002zk}, for a theory without flavors, the multiplication map allowed
reduction of all discussions of confining vacua to Coulomb vacua.   In our case with flavors, defining a formal ``multiplication index''  for $\prod U(N_i)$ theories as the greatest common divisor (GCD) of the $N_i$ and the number of flavors $2l$, we can hope to reduce some vacua of higher rank gauge theories with multiplication index $t>1$ to vacua of lower rank gauge theory  with multiplication index $t=1$.   However, there is a subtlety here that was already 
encountered in the case without flavors.   Although the multiplication map can relate different product gauge theories and give useful information concerning which classical limits can be smoothly connected, it does not specify the properties of the $\CN = 1$ vacua that are related by the map.   In particular the higher rank gauge theories involved in the map will have more vacua than the lower rank ones, and so a more refined index is useful. Such an index, called the ``confinement index'', was defined for  theories without flavors in \cite{Cachazo:2002zk} as the 
 GCD of the $N_i$ and $r_i - r_j$ where the $r_i$ label the different oblique confining vacua in each $U(N_i)$ factor.   This index 
had an attractive physical meaning in terms of the representations whose Wilson loops did not show area laws.   The same index cannot be obtained in our case because the presence of massless fundamentals ensures perfect screening -- there is never an area law and so the confinement index is always equal to 1.    However, it is clear by examining the factorized forms of the SW curve that some refinement of the ``multiplication index''  that we are using is possible.      The connection to a matrix model can probably also be used to defined a more refined index.   Recall that in the absence of flavors the correct multiplication (or confinement) index could also be obtained in terms of period integrals around all compact cycles computed from the eigenvalue density function $T(x)$ treated as a function on the complex plane.  Such a relationship should still be available in the presence of massless flavors, but the precise details remain to be elucidated in a future publication.

\section{Quadratic Tree Level Superpotential}
\label{sec:quad}
In this section we will discuss the case of a quadratic tree level
superpotential  \be \label{cubicq-W} W(\Phi)= u_2+ m u_1 \ee so
that $W'(x)=(x+m)$.  In this case the gauge group 
will necessarily contain only one factor in the 
semiclassical limit $\Lambda\rightarrow 0$. Suppose
the Seiberg-Witten curve is
$y^2=P_{N_c}(x)^2-4\Lambda^{2N_c-N_f}(x+m)^{N_f}$, then according
to our previous analysis there are various $r$'th branches, at the roots of which the SW curve 
factorizes as
$P_{N_c}(x,u_k)=(x+m)^{r} P_{N_c-r}(x)$.

In a quadratic tree level superpotential the points at the root of
the $r$'th branch that are not lifted have at least $N_c-r-1$
massless monopoles condensed.   Thus we require at least $N_c-r-1$
double roots in the factor
$P_{N_c-r}(x)^2-4\Lambda^{2N_c-N_f}(x+m)^{N_f-2r}$ of the
Seiberg-Witten curve. There is a special case, only for $r = N_f - N_c$,  where we have one
extra double root.  This is what we called the ``baryonic branch'' in analogy with $SU(N)$ theories.

In this section we will use the general methods developed in Sec.~\ref{weak} and Sec.~\ref{sec:strong} to study the $r = N_f/2$ branch and the ``baryonic'' case with $r = N_f - N_c$.   For the other branches we will work out several examples since the general solution to the relevant factorization problem is not convenient (however, see \cite{janik}).
 We will find the vacua of these theories in the strong coupling region, take classical limits, and show that these limits always contain only one gauge group factor.   As such, the example of a quadratic superpotential will not allow us to study smooth interpolation between classical theories with different product gauge theories.   Nevertheless, this is a good warmup example.

\subsection{\textbf{Some general cases}}
First we consider two cases where a general discussion is
possible.  We will rely heavily on the results present in Sec.~\ref{weak} and Sec.~\ref{sec:strong} and refer the reader to those sections of clarification of all our obscurities.

\paragraph{\underline{$\mathbf{r=\frac{N_f}{2}}$:}}  In this case there is a residual $Z_2$ symmetry
so the number of vacua should be
$\frac{1}{2}(2N_c-N_f)$ from the weak coupling analysis.   We will show this is also the case from the strong coupling point of view.  The
SW curve is
\begin{eqnarray}
y^2=(x+m)^{N_f}(P_{N_c-r}(x)^2-4\Lambda^{2N_c-2r})
\end{eqnarray}
We need $N_c-r-1$ double roots in the factor
$P_{N_c-r}(x)^2-4\Lambda^{2N_c-2r}$. This has been solved
\cite{Douglas:1995nw,Ferrari:2002jp} (see also
\cite{Balasubramanian:2002tm}).
One solution is given by the  eigenvalues are \be
\Phi=diag[-m,...,-m,x_1,...,x_{N_c-r}],~~~~~x_k=2 \Lambda \cos
{\pi (k-{1\over 2}) \over N_c-r}+z \ee From this we read out
${1\over 2}\Tr(\Phi^2)={N_c-r \over 2} [z^2+ 2\Lambda^2]+{r \over 2}m^2$
and $\Tr(\Phi)=-rm+(N_c-r)z$. Putting $u_2,u_1$ into the 
superpotential $W=u_2+mu_1$ and minimizing it we get $z=-m$. Then it is easy to see that
in the limit $\Lambda\rightarrow 0$, $x_k\rightarrow -m$.  So in the classical limit we obtain a single gauge group factor.  We expect to have $N_c-N_f/2=N_c-r$ vacua, and these
are obtained by setting $\Lambda^2 \rightarrow \Lambda^2 e^{2i\pi
k/(N_c-r)}$, all of which solve the factorization condition.

\paragraph{\underline{Baryonic branch with $\mathbf{r=N_f-N_c}$:}} In this case we require
$P_{N_c}(x)=(x+m)^rP_{N_c-r}(x)$, and the Seiberg-Witten curve is
\begin{eqnarray}
y^2=(x+m)^{2r}(P_{N_c-r}(x)^2-4\Lambda^{2N_c-N_f}(x+m)^{N_c-r})
\end{eqnarray}
We require $N_c-r$ double roots in
$P_{N_c-r}(x)^2-4\Lambda^{2N_c-N_f}(x+m)^{N_c-r}$. The only
solution is $P_{N_c-r}(x)=(x+m)^{N_c-r}+\Lambda^{2N_c-N_f}$. This
is the ``baryonic'' root invariant under the unbroken $Z_{2N_c-N_f}$
symmetry. Again, the semiclassical limit has a single gauge group factor since 
$P_{N_c}\rightarrow (x+m)^{N_c}$.

\subsection{$\mathbf{U(2)  ~{\mathrm{with}} ~2}$ \textbf{flavors}}
The only possible case that is not covered by the above general discussion is the $r=0$ branch.
We need one double root
in the Seiberg-Witten curve $y^2=P_2(x)^2-4\Lambda^2(x+m)^2$.
Suppose $P_2(x)-2\eta\Lambda(x+m)=(x+a)^2$, then
\begin{eqnarray}
y^2=(x+a)^2((x+a)^2+4\eta\Lambda(x+m))
\end{eqnarray}
We can then read out the tree level superpotential $W^{\prime}(x)=x+m=x+a$
using equation (\ref{flavor-F_2n}).   In the weak coupling analysis we expect to have $2 N_c - N_f = 
2\times 2-2=2$ vacua.   From the strong coupling perspective, we have found two solutions, $a=m$ for $\eta=\pm 1$ as expected. Again the semiclassical limit recovers a single gauge group factor:
$P_2(x)\rightarrow (x+m)^2$.

\subsection{$\mathbf{U(3)~ {\mathrm{with}}~ 2}$ \textbf{flavors}}
We discuss the $r=0$ branch since other cases have been
covered by the above general discussion. We need two double roots
in the Seiberg-Witten curve $y^2=P_3(x)^2-4\Lambda^4(x+m)^2$. 
To solve the factorization problem we write:
\begin{eqnarray} \label{10001}
P_3(x)-2\Lambda^2(x+m)=(x+m-\frac{a^3}{\Lambda^2}+a)^2(x+m-\frac{a^3}{\Lambda^2}-2a-\frac{\Lambda^2}{a})
\nonumber \\
P_3(x)+2\Lambda^2(x+m)=(x+m-\frac{a^3}{\Lambda^2}-a)^2(x+m-\frac{a^3}{\Lambda^2}+2a-\frac{\Lambda^2}{a})
\end{eqnarray}
So on the factorization locus the SW curve is
\begin{eqnarray}
y^2=(x+m-\frac{a^3}{\Lambda^2}+a)^2(x+m-\frac{a^3}{\Lambda^2}-a)^2
((x+m-\frac{a^3}{\Lambda^2}-\frac{\Lambda^2}{a}))^2-4a^2)
\end{eqnarray}
Again we read out the tree level superpotential 
$W^{\prime}(x)=x+m-\frac{a^3}{\Lambda^2}-\frac{\Lambda^2}{a}$
using equation (\ref{flavor-F_2n}), and thus find that
 $\frac{a^3}{\Lambda^2}+\frac{\Lambda^2}{a}=0$.   This has four solutions which matches the weak coupling expectation of $2N_c - N_f = 4$ vacua.  The semiclassical limit has one gauge group factor:
$P_3(x)\rightarrow (x+m)^3$.

\subsection{$\mathbf{U(3)~ {\mathrm{with}}~ 4}$ \textbf{flavors}}
Since the other cases are covered by the general discussion we focus on
the non-baryonic $r=1$ and $r=0$ branches.

\paragraph{Non-baryonic $r=1$ branch:} The addition map can be used to reduce the calculations to the non-baryonic $r=0$ branch of $U(2)$ with $2$ flavors. The number
of vacua is then $2N_c - N_f = 3\times 2-4=2$ as shown in $U(2)$ case.

\paragraph{Non-baryonic $r=0$ branch:} We solve the factorization
and find
\begin{eqnarray}
P_3(x)-2\Lambda(x+m)^2=(x+a+b)^2
(x+a-2\Lambda-2b-\frac{2(m-a)\Lambda}{b}) \nonumber \\
P_3(x)+2\Lambda(x+m)^2=(x+a-b)^2 (x+a+2\Lambda+2b-\frac{2
(m-a)\Lambda}{b})
\end{eqnarray}
with the constraint $b^2(b+\Lambda)=(m-a)^2\Lambda$. Thus we read
out the tree level superpotential
$W^{\prime}(x)=x+a-\frac{2(m-a)\Lambda}{b}$.

To count the number of vacua, we need to solve the following two equations for $a$, $b$ with fixed
$m$ :
\begin{eqnarray}
a-\frac{2(m-a)\Lambda}{b}=m \nonumber \\
b^2(b+\Lambda)=(m-a)^2\Lambda
\end{eqnarray}
There are $3$ solutions $a=m\pm 2i\Lambda$, $b=-2\Lambda$ and
$a=m$, $b=-\Lambda$. One can check that we indeed have a single gauge group factor for each solution in the classical limit $\Lambda \rightarrow 0$.
The first two solutions come in the strong coupling region from
non-baryonic roots and match the weak coupling counting of vacua arising from non-degenerate meson fields $2N_c-N_f=2\times 3-4=2$.
The third solution is a little different  -- in the weak analysis it comes from the
case with degenerate meson fields $M$. Thus in the strong analysis, this vacuum should lie on a non-baryonic submanifold in the baryonic
branch \cite{Argyres:1996eh} as discussed in Sec.~\ref{sec:strong}. 
  Indeed, for the third solution the SW curve factorizes as
\begin{eqnarray}
P_3(x)-2\Lambda^2(x+m)=(x+m-\Lambda)^2(x+m) \nonumber \\
P_3(x)+2\Lambda^2(x+m)=(x+m+\Lambda)^2(x+m)
\end{eqnarray}
This happens to match the SW curve on the  $r=1$ branch. This coincidence occurs because this vacuum lies  on the non-baryonic submanifold of the baryonic branch  which exists for  $r=N_f-N_c=1$ . 

The reason why we always obtain a single gauge group factor in the classical limit is because we require the bare squark mass $m$ to be a root of the superpotential so that we have charged massless quarks.  Because of this there was a relation between the Coulomb moduli and the $m$, which in turn led to a single gauge group factor in the classical limit.

\section{Cubic Tree Level Superpotential}
\label{sec:cubic}
In this section we will discuss the case of a cubic tree level
superpotential 
 \be \label{cubic-W} W(\Phi)= u_3+ (m+\alpha) u_2+
m \alpha u_1 \ee 
so that $W'(x)=(x+m)(x+\alpha)$.   (We will heavily use the general results in Sec.~\ref{weak} and Sec.~\ref{sec:strong}.)  In this case the gauge group will break into two  factors in the semiclassical
limit $\Lambda\rightarrow 0$.  Suppose the Seiberg-Witten curve is
$y^2=P_{N_c}(x)^2-4\Lambda^{2N_c-N_f}(x+m)^{N_f}$. Then according
to our previous analysis there are various $r$'th branches in
which  the SW  curve factorizes as 
$P_{N_c}(x,u_k)=(x+m)^rP_{N_c-r}(x)$. In the $r>0$ branches we have
massless flavors quantumly.

With a cubic tree level superpotential the points at the root of the
$r$'th branch that are not lifted have at least $N_c-r-2$ massless
monopoles condensed.  Thus we require at least $N_c-r-2$ double
roots in the factor
$P_{N_c-r}(x)^2-4\Lambda^{2N_c-N_f}(x+m)^{N_f-2r}$ of the
Seiberg-Witten curve. Suppose we have a semiclassical limit
$U(N_c)\rightarrow U(N_1)\times U(N_2)$ with $N_f$ flavors charged
under $U(N_1)$.  Then in the semiclassical limit we are also in the
$r$'th branch of the $U(N_1)$ block with $N_f$ flavors. There are
two special cases which we call ``baryonic branches'' where we have one extra double root. One is
$r=N_f-N_1$  and another is $r=N_1$.

We will
consider various examples and solve the factorization problem in the 
baryonic $r$'th branches (with $N_c-r-1$ double roots) and
non-baryonic $r$'th branches (with $N_c-r-2$ double roots).   The methodology in these two cases is as follows:
\begin{enumerate}
\item Non-baryonic $r$'th branch: The solution to the factorization problem has
two parameters, thus by solving it we are
restricted to a two-parameter subspace of the Coulomb moduli
space. We will use equation (\ref{flavor-F_2n}) to find
 the tree level superpotential and equations relating the two parameters of the Coulomb moduli space to  the roots of superpotenital  $m,\alpha$.
 Then we discuss various semiclassical limits and possible
interpolations between them.   After that we will fix
the superpotential and  count the
number of vacua to check the general formulae given in previous
section.

\item Baryonic $r$'th branch: The factorization problem has only one
parameter. Also here the gauge group breaking patterns are very
restrictive:  we must have $N_1=r$ or $N_1=N_f-r$, and for either
case the number of vacua is $N_2=N_c-N_1$. 
Because of this the possible
semiclassical limits are completely determined.  We will show how this follows from the SW curve and explore the possible smooth interpolations between baryonic vacua.  We will only determine the tree level superpotential and count the number of vacua  for one example.  This is because an additional monopole is condensed in this case and therefore (\ref{flavor-F_2n}) should be generalized  before it  can be directly applied to determine the superpotential.

\end{enumerate}

We find it useful to introduce some notation at this stage. Henceforth, we will use the notation $\widehat{U(N_1)} \times U(N_2)$ to denote a product group theory with gauge group $U(N_1) \times U(N_2)$ and flavors charged under the $U(N_1)$ factor.

\subsection{\textbf{The $\mathbf{U(2)}$ gauge group}}
For $U(2)$  we will discuss the cases with 0 and 2 flavors. We will not 
discuss an 
odd number of flavors in our paper except in the baryonic branch
 because the calculations are more difficult.  We also constrain ourselves to the case $N_f<2N_c$ so that the theory is asymptotically free and we can use the Seiberg-Witten curve safely. 

\subsubsection{$\mathbf{N_f=0}$}

The $N_f=0$ case has been discussed in \cite{Cachazo:2002zk}. The curve
is
\be \label{U2-N-0}
y^2=[(x+a)(x+b)]^2-4 \Lambda^4
\ee
From this we read out $\Phi=\diag\{-a,-b\}$ and $W'(x)=(x+a)(x+b)$,
so $a=m,~b=\alpha$. In the IR a $U(1)\times U(1)$ survives.  After minimizing the superpotential there is only one $\CN = 1$ vacuum.

\subsubsection{$\mathbf{N_f=2}$}
In this case we can have $r=0$ and $r=1$ branches. However, since
the only possible classical breaking pattern $U(2)\rightarrow U(1)\times \widehat{U(1)}$ (the $~\widehat{  }~$ denotes the flavors are charged under the $U(1)$ factor), the $r=1$ branch is a baryonic
branch (since $r=N_1$) while the $r=0$ branch is non-baryonic.

\paragraph{\underline{Baryonic $r=1$ branch:}} 
The $r=1$ branch can only be a
baryonic branch since the classical limit is $U(1)\times \widehat{U(1)}$
with $r=1=N_1$. The curve is
\be \label{U2-2-r-1}
y^2= (x+m)^2 [(x+a)^2-4 \Lambda^2]
\ee
where no extra double root is needed. Putting $\Phi=\diag\{-m,-a\}$ 
into the superpotential (\ref{cubic-W}) and minimizing it we
find that $a=\alpha$. In classical limit $P_2(x)=(x+m)(x+\alpha)$
so we get $U(2)\rightarrow U(1)\times \widehat{U(1)}$. However, since this is a baryonic branch,
the  $U(1)$ factor with flavors is actually higgsed.  We are left
only one decoupled $U(1)$ without flavors.

\paragraph{\underline{Non-baryonic $r=0$ branch:}} We do not need double roots
in  the Seiberg-Witten curve
$y^2=P_2(x)^2-4\Lambda^2(x+m)^2$. Using (\ref{flavor-F_2n})
 we read off the corresponding tree level
superpotential.   (Notice in this case we need to take into account
the modification of the formula with flavors since $N_f\geq
2N_c-2$.) 
$$F_4(x)+4\Lambda^2(x+m)^2=P_2(x)^2=W^{\prime}(x)^2+{\cal
O}(x)~.$$ So $P_2(x)=W^{\prime}(x)=(x+\alpha)(x+m)$,
 thus the semiclassical limit is
$U(2)\rightarrow U(1)\times \widehat{U(1)}$. To count the number of vacua, we see for fixed $m$
and $\alpha$ there is only one solution for $P_2(x)$, so there is
only one vacuum.

Notice that putting
$P_2(x)=(x+\alpha)(x+m)$ into the SW curve we find 
\be
y^2=[(x+\alpha)(x+m)]^2-4\Lambda^2(x+m)^2=(x+m)^2[(x+a)^2-4 \Lambda^2]
\ee
which has the same form as the SW curve as in the $r=1$ branch. 
In fact we 
encountered a similar  phenomenon for a quadratic superpotential, in a 
$U(3)$ theory with four flavors in the $r=0$ branch.  There the explanation was that the non-baryonic $r=0$ branch lies inside the baryonic $r=1$ branch.   The same thing applies here, and we should count this vaccum as a solution for each of these branches.

\subsection{\textbf{The $\mathbf{U(3)}$ gauge group}}
\label{sec:u3}
Things get more interesting with a $U(3)$ gauge group since there are two different classical breaking patterns:   $U(3)\rightarrow U(1)\times U(2)$  with massless flavors in either the $U(1)$ or the $U(2)$ factor.   We will consider cases with $N_f=0,2,4$.   The number of vacua arising in each case is 
summarized in Table~\ref{U3-summation}.

\TABLE[h]
{\begin{tabular}{|c|c|c|c|} \hline
$N_f$ & $r$  &   $U(1) \times \widehat{U(2)}$  & $\widehat{U(1)}\times U(2)$ \\  \hline
0 & 0  &  2  & 2  \\  \hline
$2$ & 0  &  $2/(1)$  &   $2$  \\ \hline
& 1  &  1  & (2)   \\ \hline
$4$ & 0 &  1 & 2 \\ \hline
 & 1 &  1 & (2)   \\ \hline
& 2 & (1) &  \\ \hline
\end{tabular}
\caption{We remind the reader of our notation.  By  $U(1) \times \widehat{U(2)}$ we mean the classical limit 
$U(3)\rightarrow U(2)\times U(1)$ with flavors charged under
$U(2)$.  The empty block means that there is no vacua for this case.
For example we can not have  an $r=2$ branch for 
$U(3)\rightarrow U(2)\times \widehat{U(1)}$. Also we use $n$ and $(n)$ to distinguish the non-baryonic 
and baryonic vacua. If for any given $r$ both kinds of branches exist, we use $n/(m)$ for
them.\label{U3-summation}
}}

\subsubsection{$\mathbf{N_f=0}$}
This case has been studied in \cite{Cachazo:2002zk}. For our cubic superpotential
there are four solutions which are divided into two groups corresponding to which root of the superpotential the $U(2)$ is located at. Each group has two vacua corresponding to the two confining vacua of $U(2)$.

\subsubsection{$\mathbf{N_f=2}$}
Here, we can have   $r=0$ and $r=1$ branches. For $U(3)\rightarrow
\widehat{U(1)}\times U(2)$,  the $r=1$ branch is
baryonic ($r=N_c$) while $r=0$ branch is non-baryonic. For the breaking
pattern $U(3)\rightarrow
\widehat{U(2)}\times U(1)$, non-baryonic branches  exist for
both $r=0$ and $r=1$ while a baryonic branch exists only for $r=0$ branch.

\paragraph{\underline{Baryonic $r=1$ branch:}}
The only possible gauge group breaking pattern is $U(3)\rightarrow
\widehat{U(1)}\times U(2)$.  The Seiberg-Witten curve is
$y^2=(x+m)^2(P_2(x)^2-4\Lambda^4)$. We need one double root in
$P_2(x)^2-4\Lambda^4$. Suppose
\be \label{U3-Nf2-r-1}
P_2(x)+2 \eta \Lambda^2= (x+a)^2\Rightarrow u_1=-2a,~~u_2=a^2+\eta \Lambda^2 \, .
\ee
Using $u_3=u_1 u_2-{1\over 6} u_1^3$, and minimizing the superpotential, we get
$ a^2-\alpha a +\eta \Lambda^2=0$. From this we have a solution
$a=(\alpha\pm \sqrt{\alpha^2-4 \eta \Lambda^2})/2$. However, to get two gauge group factors in the classical limit $\Lambda\rightarrow 0$ we need to choose $+$ sign so that 
$a\rightarrow \alpha$. Thus we get two solutions labeled by the choice of  $\eta=\pm 1$.
In the weak coupling limit, the $U(1)$ factor with two flavors has one vacuum while the confining $U(2)$ has two.  So there are $1 \times 2 = 2$ weak coupling vacua matching the strong coupling result.

\paragraph{\underline{Non-baryonic $r=1$ branch:}} In this case we do not need a double
root. The Seiberg-Witten curve $y^2=P_3(x)^2-4\Lambda^4(x+m)^2$ is factorized as 
\begin{eqnarray}
y^2=(x+m)^2(P_2(x)^2-4\Lambda^4) \, .
\end{eqnarray}
So $W^{\prime}(x)=P_2(x)=(x+m)(x+\alpha)$ by equation (\ref{flavor-F_2n}). 
The semiclassical limit gives
$U(3)\rightarrow U(1)\times \widehat{U(2)}$. To count the number of vacua, observe that we have only $1$ solution
for $P_2(x)$ with fixed $m$ and $\alpha$. This is consistent with the weak coupling analysis since
in this case with $r=\frac{M_i}{2}$ we expect $\frac{1}{2} (2N_i - M_i) = \frac{1}{2}(2\times
2-2)=1$ vacua.

\paragraph{\underline{Baryonic $r=0$ branch:}} We need two double
roots in the Seiberg-Witten curve
$y^2=P_3(x)^2-4\Lambda^4(x+m)^2$.   Solving the factorization
problem gives
\begin{eqnarray} \label{10002}
P_3(x)-2\Lambda^2(x+m)=(x+m-\frac{a^3}{\Lambda^2}+a)^2(x+m-\frac{a^3}{\Lambda^2}-2a-\frac{\Lambda^2}{a})
\nonumber \\
P_3(x)+2\Lambda^2(x+m)=(x+m-\frac{a^3}{\Lambda^2}-a)^2(x+m-\frac{a^3}{\Lambda^2}+2a-\frac{\Lambda^2}{a})
\end{eqnarray}
For a non-singular semiclassical limit we must have
$a\sim \Lambda^p$ with $\frac{2}{3}\leq p\leq 2$. But when $p< 2$
this limit leads to a single gauge group factor: $P_3(x)\rightarrow
(x+m-\frac{a^3}{\Lambda^2})^3$. So the only non-singular
semiclassical limit with two gauge group factors arises by keeping $v=-\frac{\Lambda^2}{a}$ fixed.
Then $P_3(x)\rightarrow (x+m)^2(x+m+v)$.

To count the number of vacua we first fix $m$ and
$\alpha$ in the tree level superpotential and find the number of
$a$ which extremize it. Since this is a baryonic branch where we have
one extra monopole, it is not straightforward to use the equation (\ref{flavor-F_2n}).
So we will do the extremization directly.
First from (\ref{10002}) and the formula
$P_3(x)=x^3-u_1x^2+(\frac{u_1^2}{2}-u_2)-\frac{u_1^3}{6}+u_1u_2-u_3$
we find $u_1$, $u_2$, $u_3$. We then minimize the tree level
superpotential $W=u_3+(m+\alpha)u_2+m\alpha u_1$. We find
\begin{eqnarray}
\frac{\partial{W(m,\alpha,a)}}{\partial
a}=\frac{(9a^4-\Lambda^4)(a^8+(\alpha-m)\Lambda^2a^5
+4\Lambda^4a^4+(\alpha-m)\Lambda^6a+\Lambda^8)}{a^4\Lambda^6}
\end{eqnarray}
Vacua arising from the solution $9a^4-\Lambda^4=0$ are not interesting to us because they cannot be brought to a weak coupling region by varying parameters of the superpotential.
Discarding these solutions we can analyze the asymptotic behavior of the remaining eight
roots when $\Lambda\rightarrow 0$
\begin{equation} \label{100021}
a^8+(\alpha-m)\Lambda^2a^5
+4\Lambda^4a^4+(\alpha-m)\Lambda^6a+\Lambda^8=0
\end{equation}
There
are three roots $a^3\sim(m-\alpha)\Lambda^2$ for which the first
and the second terms in (\ref{100021}) dominate as $\Lambda \rightarrow 0$.
There is one root
$a\sim\frac{\Lambda^2}{m-\alpha}$ for which the fourth and the
fifth terms in (\ref{100021}) dominate.   Finally there are 
four roots $a^4\sim -\Lambda^4$ for which 
the second and the fourth terms in (\ref{100021}) dominate.   

As we have already analyzed, the four roots $a^4\sim -\Lambda^4$ have semiclassical limits with a single gauge group factor: $P_3(x)\rightarrow (x+m)^3$.   The three roots
$a^3\sim(m-\alpha)\Lambda^2$ also have a single gauge factor in the semiclassical
limit: $P_3(x)\rightarrow (x+\alpha)^3$.\footnote{However, we will see later by using the addition map that this case will explain the physics of  $U(4)$ theory with $N_f=4$ in the $r=1$ branch.}
The single root
$a\sim\frac{\Lambda^2}{m-\alpha}$ has the semiclassical limit
$P_3(x)\rightarrow (x+m)^2(x+\alpha)$.  This the solution with two gauge group factors that we are looking for.   The single vacuum matches the counting in the weak coupling region.

\paragraph{\underline{Non-baryonic $r=0$ branch:}} The Seiberg-Witten curve
$y^2=P_3(x)^2-4\Lambda^4(x+m)^2$ must have a double root. Suppose
$P_3(x)-2\eta\Lambda^2(x+m)=(x+a_1)^2(x+a_2)$. Then the Seiberg-Witten
curve is
\begin{eqnarray}
y^2&=&P_3(x)^2-4\Lambda^4(x+m)^2=(x+a_1)^2(x+a_2)((x+a_1)^2(x+a_2)+4\eta\Lambda^2(x+m))
\nonumber \\ &=&(x+a_1)^2(((x+a_1)(x+a_2)+2\eta\Lambda^2)^2+{\cal
O}(x))
\end{eqnarray}
So $W^{\prime}(x)=(x+a_1)(x+a_2)+2\eta\Lambda^2$.    There are two
solutions for given $(a_1,a_2)$: $m=\frac{1}{2}(a_1+a_2\pm\sqrt{(a_1-a_2)^2-8\eta\Lambda^2})$,
$\alpha=\frac{1}{2}(a_1+a_2\mp\sqrt{(a_1-a_2)^2-8\eta\Lambda^2})$.
 In the semiclassical
limit, one solution gives the breaking pattern $U(3)\rightarrow \widehat{U(1)}\times U(2)$  and the other solution gives 
$U(3)\rightarrow U(1)\times \widehat{U(2)}$. These two breaking pattern do not interpolate between
each other because they arise as different solutions for the vacua.

To count the number of vacua, we need to fix $m$ and $\alpha$ and
find the number of solutions for $a_1$ and $a_2$. For each
$\eta=\pm 1$ there are two solutions. In the semiclassical limit they
become $a_1\rightarrow m$, $a_2\rightarrow \alpha$ and
$U(3)\rightarrow U(1)\times \widehat{U(2)}$, or $a_1\rightarrow \alpha$, $a_2\rightarrow m$ and
$U(3)\rightarrow \widehat{U(1)} \times U(2)$.    The count of vacua is consistent with the weak coupling limit as follows.
For flavors under the $U(2)$,  the factor $U(1)$ has one vacuum and the $U(2)$ with $N_f =2$ in the $r=0$ branch has $2N_c - N_f = 2$ vacua giving a total of $1 \times 2 = 2$ vacua.     For flavors under $U(1)$, the confining $U(2)$ factor leads to 2 vacua.

\paragraph{\underline{Addition map:}}  We can also explain these results from the point of view of the addition map that we introduced.   We have claimed that 
$U(3)$ with $N_f=2$ at $r=1$ branch can be reduced to $U(2)$ with $N_f=0$ at
$r=0$ branch. From our previous discussion in $U(2)$ section, the later has only 
one solution for $U(2)\rightarrow U(1)\times U(1)$ .
Using the addition map to go from $U(2)$ to
$U(3)$ we get only one solution for 
$U(3)\rightarrow \widehat{U(2)}\times U(1)$. 
Note that the addition map does not
recover the case with flavors  charged under $U(1)$.    This is because 
$U(1)$ with $N_f=2$ at $r=1$ will formally reduce to $U(0)$ with $N_f=0$ in the $r=0$ branch
which does not exist. In fact, since the case with flavors in the $U(1)$
 is in the baryonic branch we should
not expect to recover it from the non-baryonic branch in $U(2)$. 
This example demonstrates  us that the general 
addition map is very useful, but does not cover all cases.

\subsubsection{ $\mathbf{N_f=3}$}
It is difficult to analyze the case with an odd number of flavors and we only study the baryonic branch where the analysis simplifies.

\paragraph{\underline{Baryonic $r=1$ branch:}}
Two semiclassical limits $U(3)\rightarrow \widehat{U(1)}\times U(2)$,  and $U(3)\rightarrow U(1)\times \widehat{U(2)}$  are possible.
As we will demonstrate shortly, we can smoothly interpolate between these classical theories by passing through the strong coupling region.

Suppose $P_3(x)=(x+m)P_2(x)$, then the Seiberg-Witten curve is
\begin{equation}
y^2=P_3(x)^2-4\Lambda^3 (x+m)^3=(x+m)^2(P_2(x)^2-4\Lambda^3 (x+m)) \, .
\end{equation}
We require $P_2(x)^2-4\Lambda^3(x+m)$ to have a double root. Suppose
\begin{eqnarray}
P_2(x)^2-4\Lambda^3(x+m)&=&((x+c)^2+a(x+c)+b)^2-4\Lambda^3(x+m)\nonumber
\\ &=&(x+c)^2((x+c)^2+2v(x+c)+u)
\end{eqnarray}
We equate the coefficients of the powers of $x$ and find $4$
equations for $a$, $b$, $c$, $u$, $v$. Eliminating $a$ and $b$ we
find
\begin{eqnarray} \label{uv}
u-v^2&=&\pm 4((m-c)\Lambda^3)^{\frac{1}{2}} \nonumber \\
v&=&\pm (\frac{\Lambda^3}{m-c})^{\frac{1}{2}}
\end{eqnarray}
We can take two different semiclassical limits 
\begin{enumerate}
\item $\Lambda\rightarrow 0$ and $c-m$ fixed and finite. Then
according to (\ref{uv}) $u\rightarrow 0$, $v\rightarrow 0$,
$P_2(x)\rightarrow (x+c)^2$. So $P_3(x)\rightarrow (x+m)(x+c)^2$,
$U(3)\rightarrow \widehat{U(1)}\times U(2)$.

\item $\Lambda\rightarrow 0$, $c\rightarrow m$ and
$v=\pm (\frac{\Lambda^3}{m-c})^{\frac{1}{2}}$ fixed. Then
$u\rightarrow v^2$, $P_2(x)\rightarrow (x+m)(x+m+v)$. So
$P_3(x)\rightarrow (x+m)^2(x+m+v)$, $U(3)\rightarrow U(1)\times
\widehat{U(2)}$.
\end{enumerate}
Thus we indeed find two semiclassical limits which are smoothly connected to each other.

Both semiclassical limits are in baryonic branches, but the vacua arise in different ways.
For flavors charged under $U(1)$, we are in the $r=N_c$ branch
while for flavors charged under $U(2)$, we are in the
$r=N_f-N_c$ branch.   However, in both cases $r=1$ and this is why there is possibility of smooth interpolation.   In both cases one of the gauge group factors (either $U(1)$ or $U(2)$)  is totally higgsed, leaving a single decoupled $U(1)$ from the other factor.  (When the $U(1)$ is Higgsed $SU(2) \subset U(2)$ confines.)   Thus the low energy
theory is identical. Generically, for this to happen with our cubic superpotential
we need $N_1=r$ or $N_1=N_f-r$.   We have seen one example here and will see another we we consider $U(4)$ with four flavors in the $r=1$ branch.

\subsubsection{${\mathbf {N_f=4}}$}
In this case we can have $r=0,1,2$ branches. The $r=2$ branch exists
only for flavors charged under $U(2)$ and is a baryonic branch.
The $r=1$ branch is non-baryonic if the flavors charged under $U(2)$
and is baryonic if  flavors are charged under $U(1)$. The $r=0$ branch is 
non-baryonic for both breaking patterns.

\paragraph{\underline{Baryonic $r=2$ branch:}} The
$r=2$ branch can only be a baryonic branch. By the addition map this reduces to a 
baryonic $r=1$ branch in $U(2)$ with two flavors that have already studied. It
is obvious that the only semiclassical limit is $U(3)\rightarrow
U(1) \times \widehat{U(2)}$.

\paragraph{\underline{Baryonic $r=1$ branch:}}
The curve is
\be
y^2=(x+m)^2[P_2(x)^2-4\Lambda^2 (x+m)^2]
\ee
Requiring one double root gives\footnote{We have put $m=0$ to make the computation simpler.}
\be \label{U2-Nf2-r-00}
P_2(x)-2 \eta \Lambda x= (x+a)^2 \, .
\ee
On this branch the Coulomb moduli are therefore
\be
u_1=-2(a+\eta \Lambda),~~u_2=a^2+ 4 \eta \Lambda a+ 2 \Lambda^2 
\ee
Actually we are just looking at the $U(2)$ arising from the part of the SW curve inside the square brackets.   The $u_n$ for the full $U(3)$ are given by shifting the above by $(-m)^n/n$.
For the $U(2)$ factor we also have $u_3=u_1 u_2-{1\over 6} u_1^3$.     Putting everything
into the superpotential which now becomes $W(\Phi)=u_3+\alpha u_2$ and minimizing it
we get
\be \label{U2-r=0}
a={1\over 2}[ \alpha-6 \eta \Lambda \pm \sqrt{\alpha^2-4 \eta \Lambda \alpha+20\Lambda^2}]
\ee
To have two classical gauge group factors we need to take $+$ sign which 
gives us $U(2)\times \widehat{U(1)}$.
We have two solutions for $\eta=\pm 1$.  In the weak coupling region the corresponding two vacua arise from confinement of $SU(2) \subset U(2)$.

\paragraph{\underline{Non-baryonic $r=1$ branch:}} 
This case reduces by the addition map to the non-baryonic $r=0$ branch in $U(2)$ with two flavors.
The semiclassical limit is $P_3(x)\rightarrow (x+m)^2(x+\alpha)$ and
$U(3)\rightarrow U(1)\times \widehat{U(2)}$. As in the $U(2)$ case, the curve of this $r=1$ branch looks like the curve for an $r=2$ branch, again because we are considering a non-baryonic branch embedded within a baryonic one.

\paragraph{\underline{Non-baryonic $r=0$ branch:}} The Seiberg-Witten curve
$y^2=P_3(x)^2-4\Lambda^2(x+m)^4$ has a double root. Suppose
$P_3(x)-2\eta\Lambda (x+m)^2=(x+a_1)^2(x+a_2)$. The Seiberg-Witten
curve is
\begin{eqnarray}
y^2 & = & P_3(x)^2-4\Lambda^2(x+m)^4\nonumber \\
&= &(x+a_1)^2(x+a_2)((x+a_1)^2(x+a_2)+4\eta\Lambda(x+m)^2)=(x+a_1)^2F(x)
\end{eqnarray}
So $F(x)=(x+a_1)^2(x+a_2)^2+4\eta\Lambda(x+m)^2(x+a_2)$, and
\begin{eqnarray}
F(x)+\frac{4\Lambda^2(x+m)^4}{(x+a_1)^2} &=&F(x)+4\Lambda^2x^2+{\cal
O}(x)\nonumber \\
& = & ((x+a_1)(x+a_2)+2\eta\Lambda (x-a_1+2m))^2+{\cal O}(x)
\end{eqnarray}
from which we read out
 $W^{\prime}(x)=(x+a_1)(x+a_2)+2\eta\Lambda (x-a_1+2m)$. There
are two solutions: $m=a_1$, $\alpha=a_2+2\eta\Lambda$ or
$m=a_2-2\eta\Lambda$, $\alpha=a_1+4\eta\Lambda $. The first
solution coincides with an $r=2$ branch and leads to the classical limit $U(3) \rightarrow U(1) \times \widehat{U(2)}$ 
 while the second solution becomes
$U(3)\rightarrow \widehat{U(1)}\times U(2)$.

To count the number of vacua, we need to fix $m$ and $\alpha$ and
find the number of solutions for $a_1$ and $a_2$. In the first
solution $a_1=m$, $a_2=\alpha-2\eta\Lambda$ we have
\begin{eqnarray}
P_3(x)-2\eta\Lambda(x+m)^2=(x+m)^2(x+\alpha-2\eta\Lambda)
\nonumber \\
P_3(x)+2\eta\Lambda(x+m)^2=(x+m)^2(x+\alpha+2\eta\Lambda)
\end{eqnarray}
The solution is the same for $\eta=\pm 1$, so there is actually
only one solution with a semiclassical limit $U(3)\rightarrow U(1)\times \widehat{U(2)}$.   According to the weak coupling analysis, we should
expect two vacua from the $U(2)$ factor with four flavors. 
However, since the curve coincides with that on the $r=2$ branch, we should count the weak coupling  vacua for with $r=2$. This counting gives one vacuum, consistent with the above analysis. 
For the second solution $a_1=\alpha-4\eta\Lambda$ and
$a_2=m+2\eta\Lambda$, the choices $\eta=\pm 1$ are different;  so we have two 
vacua for $U(3)\rightarrow \widehat{U(1)}\times U(2)$.  The two vacua come from the confininement in the $U(2)$ factor.

Before ending this subsection, note a special fact concerning
flavors charged under the $U(2)$ factor: all $r=0,1,2$ branches have the 
same curve as in $r=2$ branch after the minimization of the superpotential.   We we later use this fact when use the addition map to study the $U(4)$ with six flavors charged under a $U(3)$ factor.

\subsection{\textbf{The $\mathbf{U(4)}$ gauge group}}
For the $U(4)$ gauge group we will discuss $N_f=0,2,4,6$ flavors. There are three
breaking patterns $U(4)\rightarrow  U(2)\times \widehat{U(2)}$, $U(4)\rightarrow  \widehat{U(1)}\times U(3)$
 and $U(4)\rightarrow  U(1)\times \widehat{U(3)}$.   In this case we will observe smooth interpolations between
classical limits with different product gauge groups.   We summarize the number of
vacua for various $N_f,r$ and the possible interpolations in the following table. 

\TABLE[h]
{{\small
\begin{tabular}{|c|c|c|c|c|c|} \hline
$N_f$ & $r$  &   $\widehat{U(2)} \times U(2)$  &  $\widehat{U(1)} \times U(3)$  &  $\widehat{U(3)} \times U(1)$ 
& connection\\  \hline
0 & 0  &  2+2  & 3 & 3 & 
yes
\\  \hline
$2$ & 0  &  2+2/(2)  &  3 & 4 & 
yes
\\ \hline
& 1  &  2 & (3)  & 2 & no   \\ \hline
$4$ & 0 &  2+2 & 3 & 3  & 
yes
\\ \hline
 & 1 &  2 & (3) & 2/(1) & no  \\ \hline
& 2 & (2) & & 1 & no \\ \hline
$6$ & 0 & 2+2 & 3 & 1 & no \\ \hline
& 1 & 2 & (3) & 1 & no \\ \hline
& 2 & (2) & & 1 & no \\ \hline
& 3 & & & (1) & no \\ \hline
\end{tabular}
\caption{We have indicated the number of non-baryonic and baryonic vacua as ``n'' and ``(n)'' respectively.  In cases where there are smooth interpolations between classical limits with different gauge symmetry breaking patterns we have indicated a ``yes'' in the last column.\label{U4-summation}
}
}}
The details of the table are explained below.

\subsubsection{${\mathbf{N_f=0}}$}
The flavorless case has been studied in \cite{Cachazo:2002zk}. As discussed in Sec~\ref{gen}, we can label the various branches by integers $(s_+,s_-)$ which indicate the number of double roots in different factors of the SW curve. 
There are four vacua for
$U(4)\rightarrow U(2)\times U(2)$,  two of which are confining and  lie on the  $(2,0)$ and $(0,2)$ branches, and 
two of which are Coulomb vacua and lie in the $(1,1)$ branch. We denote these two vacua in the two different branches as
$2+2$ in Table~\ref{U4-summation}.
There are six vacua for $U(4)\rightarrow U(1)\times U(3)$ arising from two different distributions of the roots of the superpotential in the factors of the SW curve, and the confinement of the $U(3)$ factor which gives three vacua in each case.
Furthermore, the Coulomb vacua in the semiclassical limit with $U(2) \times U(2)$ symmetry are continuously conencted to the $U(1) \times U(3)$ vacua.

\subsubsection{$\mathbf{N_f=2}$}
In this case we can have $r=1$ and $r=0$ branches. For the breaking  
pattern with flavors charged under $U(2)$, the $r=1$ branch is non-baryonic
while $r=0$ branch can be non-baryonic or baryonic. 
For the pattern with flavors under $U(1)$, the $r=1$ branch is baryonic
while the $r=0$ branch is non-baryonic. For the pattern with 
flavors under $U(3)$, both $r=1$ and $r=0$ branches must be
non-baryonic. 
All these discussions can be summarized by the table
\TABLE[h] 
{\begin{tabular}{|c|c|c|c|c|} \hline
 & $U(2)\times\widehat{U(2)}$ & $\widehat{U(1)} \times U(3)$  & ${\widehat{U(3)}} \times U(1)$ & connection\\  \hline
$r=1$ & N & B & N  & no\\  \hline
$r=0$ & B/N & N & N & all three connected\\  \hline
\end{tabular}
\caption{We remind the reader that $U(N_1)\times \widehat{U(N_2)}$ represents a semiclassical limit with gauge group $U(N_1) \times U(N_2)$ and flavors charged under $U(N_2)$.  $B$ labels  baryonic branches and $N$ labels non-baryonic branches. The last column indicates if vacua in these classical limits can be smoothly transformed into one another.\label{U4-N_f=2-table}}}

\paragraph{\underline{Baryonic $r=1$ branch:}}
The only semiclassical limit we expect is $U(4)\rightarrow
\widehat{U(1)}\times U(3)$. Indeed,
suppose $P_4(x)=(x+m)P_3(x)$ and that $P_3(x)^2-4\Lambda^6$ has two
double roots.   Using the addition map  this is exactly the same problem we faced in the
baryonic $r=0$ branch of $U(3)$ with no flavors. We have shown
there that we can find three solutions with the semiclassical limit 
$P_4(x)\rightarrow (x+m)(x+\alpha)^3$.   In the weak coupling analysis these three vacua come
from confinement in $SU(3) \subset U(3)$. 

\paragraph{\underline{Non-baryonic $r=1$ branch:}}
 In this case $P_4(x)=(x+m)P_3(x)$ and $P_3(x)^2-4\Lambda^6$ has one
double root. Suppose $P_3(x)-2\eta\Lambda^3=(x+a_1)^2(x+a_2)$,
then
\begin{eqnarray}
y^2=(x+a_1)^2((x+a_1)^2(x+a_2)^2+4\eta\Lambda^3(x+a_2)) \, .
\end{eqnarray}
So we read off the tree level superpotential
$W^{\prime}(x)=(x+a_1)(x+a_2)=(x+m)(x+\alpha)$, which gives two
solutions $m=a_1$, $\alpha=a_2$ or $m=a_2$, $\alpha=a_1$. The
corresponding semiclassical limits are $P_4(x)\rightarrow
(x+m)^3(x+\alpha)$ or $P_4(x)\rightarrow(x+m)^2(x+\alpha)^2$. To
count the number of vacua, note that for each $\eta=\pm 1$ and fixed
$m$ and $\alpha$ there are two solutions for $P_4(x)$, so the
number of vacua for these two semiclassical limits are both two.
This is consistent with the weak coupling analysis since $r=\frac{N_f}{2}$ and thus  the number of
vacua is given by  ${1 \over 2} (2N_c - N_f) = \frac{1}{2}(2\times 3-2)\times 1=2$ for $P_4(x)\rightarrow
(x+m)^3(x+\alpha)$, and also ${1 \over 2} (2N_c - N_f)  \times 2= \frac{1}{2}(2\times 2-2) \times 2=2$ for $P_4(x)\rightarrow(x+m)^2(x+\alpha)^2$.  The final factor of two in the latter case comes from the confinement of $SU(2)$. These results can also be easily derived by the addition map which reduces
the problem to the non-baryonic $r=0$ branch of $U(3)$ without
flavors.

There is another interesting phenomenon happening here.  The SW curves of above two
solutions are
\bea \label{U4r-1-1}
y^2 &= & (x+m)^3 (x+\alpha)^2 [(x+\alpha)^2 (x+m)-4 \eta\Lambda^3]\\
y^2 &= & (x+m)^4 (x+\alpha) [(x+m)^2 (x+\alpha)-4 \eta\Lambda^3]
\label{U4r-1-2}
\eea
Notice the  factors $(x+m)^3$ and  $(x+m)^4$. They indicate the presence of nontrivial conformal fixed points in the $\CN =2$ theory.
\cite{Argyres:1995jj,Argyres:1995xn,Eguchi:1996vu}.\footnote{Specifically,
these points are in the class 4 of the classification in \cite{Eguchi:1996vu}
with $p=1$ for (\ref{U4r-1-1}) and class 4 with $p=2$ for (\ref{U4r-1-2}).}
 In fact, the appearance of
such points is a general phenomenon in our framework. 
In the flavorless case, the Seiberg-Witten curve is factorized as the product of $H_{N-n}(x)^2$
and $F_{2n}$ at points which survive the $\CN =1$ deformation. The $F_{2n}$ factor is parametrized by $n$ variables. It is 
easy to see that some  values of these $n$ variables will lead to a common factor with $H_{N-n}(x)$. In the flavorless case
we can adjust the 
superpotential $W(\Phi)$ freely, and so we can avoid such a special point.
Our treatment of massless flavors gives a different story. Since we now
require the squark mass to be a root of the superpotential,  
we have less freedom in the choice of  $W(\Phi)$ and the non-trivial superconformal fixed points
are generically the locations in the $\CN =2$ moduli space that are not lifted by the $\CN = 1$ deformation.

\paragraph{\underline{Baryonic $r=0$ branch:}}
On general grounds the only semiclassical limit with two gauge group factos that 
we expect is $U(4)\rightarrow
U(2)\times \widehat{U(2)}$.  The Seiberg-Witten curve
$y^2=P_4(x)^2-4\Lambda^6(x+m)^2$ must  have three double roots. We solve
the factorization problem and find
\begin{eqnarray}
P_4(x)-2\eta\Lambda^3(x+m)&=&(x+a)^2((x+a)^2+2(b+c)(x+a)+b^2+c^2+4bc)
\nonumber \\
P_4(x)+2\eta\Lambda^3(x+m)&=&(x+a+b)^2(x+a+c)^2
\end{eqnarray}
with the constraints
\begin{eqnarray}
bc=\pm 2\eta ((m-a)\Lambda^3)^{\frac{1}{2}}
\nonumber \\
b+c=\pm (\frac{\Lambda^3}{m-a})^{\frac{1}{2}}
\end{eqnarray}
There are two semiclassical limits
\begin{enumerate}
\item $\Lambda\rightarrow 0$, with $a-m$ fixed and finite.  Then from
the constraints we must have both $b,c \rightarrow 0$, so that
$P_4(x)\rightarrow (x+a)^4$.  This gives a semiclassical limit with one gauge group factor.

\item $\Lambda\rightarrow 0$, $a\rightarrow m$.  Then, from
the constraints, only one of $b,c$ goes to zero if we keep 
keeping $\frac{\Lambda^3}{m-a}$ fixed. Suppose $b$ is finite\footnote{Since $b$ and $c$ appear symmetrically, when $c$ is kept finite, we get the same solution.}, then
$P_4(x)\rightarrow (x+m)^2(x+m+b)^2$.  This leads to a semiclassical limit with two gauge group factors.
\end{enumerate}
So we reproduce our expectation that the only semiclassical limit with two gauge group factors gives a $\widehat{U(2)} \times U(2)$.   However, both semiclassical limits are smooth.   So here we are seeing a smooth interpolation between vacua of a $U(4)$ theory with no flavors and a $U(2) \times \widehat{U(2)}$ theory.   Physically, in the semiclassical $U(4)$ limit the flavors are classically massive, so that the $SU(4) \subset U(4)$ confines leaving a decoupled $U(1)$.   In the $U(2) \times \widehat{U(2)}$ limit,  $\widehat{U(2)}$ is in the baryonic branch and is completely Higgsed.  Meanwhile, in the other $U(2)$ the $SU(2)$ confines, leaving a decoupled $U(1)$.  Thus the low energy physics is the same in both limits.  
We already observed a similar interpolation in the baryonic $r=0$ branch of $U(3)$
with two flavors but we did not focus on this point there.
Note that  an interpolation between vacua of theories with different numbers of gauge group factors could not have happened in a non-baryonic branch since the number of low energy $U(1)$s in these vacua would necessarily be different.\footnote{Below, when $N_f=4$ or $6$, we should not expect to find a similar interpolation between vacua of a theory with different number of group factors. This is because in those cases, the we do not have two different types of baryonic branches (with $r=N_c$ and $r=N_f-N_c$). }

By  the addition map, we can generalize this  calculation to the case of
$U(5)$ theory with $4$ flavors in the baryonic $r=1$ branch. We
have two cases: $U(5)\rightarrow \widehat{U(1)}\times U(4)$ and $U(5)\rightarrow U(2)\times
\widehat{U(3)}$. The calculation done
here tells us that for baryonic branch these two classical limits in $U(5)$ 
are smoothly connected.

\paragraph{\underline{Non-baryonic $r=0$ branch in $(2,0)/(0,2)$ 
distributions:}} We need $2$
double roots in the curve for this branch and distribute them as
$(s_+,s_-) = (2,0), (0,2), (1,1)$ in the factors of the SW curve. For the distribution
$(2,0)/(0,2)$, we have $P_4(x)-2\eta\Lambda^3
(x+m)=(x+a_1)^2(x+a_2)^2$. Then
\begin{eqnarray}
F(x)=(x+a_1)^2(x+a_2)^2+4\eta\Lambda^3 (x+m)
\end{eqnarray}
So $W^{\prime}(x)=(x+a_1)(x+a_2)$. There are two solutions:
$m=a_1$, $\alpha=a_2$ or $m=a_2$, $\alpha=a_1$. The semiclaasical
limit gives $U(4)\rightarrow U(2)\times \widehat{U(2)}$.

To count the number of vacua, we fix $m$ and $\alpha$ and solve for 
$a_1$ and $a_2$. Since $a_1$ and $a_2$ are symmetric, we have only
one solution for each $\eta=\pm 1$.  From the weak coupling point of view, the breaking pattern 
$U(2)\times \widehat{U(2)}$ in the $r=0$ non-baryonic branch will have four
vacua: $2(2N_c - N_f) = (2\times 2-2)\times 2=4$.   Two of these have been found above
in  the $(2,0)/(0,2)$ distributions of double roots.  Another two will be found
in the  $(1,1)$ distribution described below.

\paragraph {\underline{Non-baryonic $r=0$  branch in 
the $(1,1)$ distribution:}}  We need $2$ double roots.
We use the freedom of translation in $x$
to tune the two double roots to be $a$ and  $-a$. The general case
can be recovered by shifting: $x\rightarrow x+b$, $m\rightarrow
m-b$. The factorization is
\begin{eqnarray}
P_4(x)+2\Lambda^3(x+m)=(x-a)^2((x+a)^2+\frac{\Lambda^3}{a^3}(mx+2ma-a^2))
\nonumber \\
P_4(x)-2\Lambda^3(x+m)=(x+a)^2((x-a)^2+\frac{\Lambda^3}{a^3}(mx-2ma-a^2))
\end{eqnarray}
From this we read out $F_{2n}(x)$ as (now $n=2$)
\begin{eqnarray}
F_4(x) &=&
((x+a)^2+\frac{\Lambda^3}{a^3}(mx+2ma-a^2))((x-a)^2+\frac{\Lambda^3}{a^3}(mx-2ma-a^2))
\nonumber \\ &=&
(x^2+\frac{m\Lambda^3}{a^3}x-a^2(1+\frac{\Lambda^3}{a^3}))^2+{\cal
O}(x)
\end{eqnarray}
So $W^{\prime}(x)$ is
\begin{equation}\label{W42}
W^{\prime}(x)=x^2+\frac{m\Lambda^3}{a^3}x-a^2(1+\frac{\Lambda^3}{a^3})
\end{equation}
There are two solutions for $m$ by setting $x=-m$ in (\ref{W42}).
(Recall that we require that $x=-m$ is one solution of $W'(x) = 0$.)  We find
\begin{equation} \label{agp}
m=\pm a\sqrt{\frac{a^3+\Lambda^3}{a^3-\Lambda^3}}
\end{equation}

These are the superpotentials that leave this point in the Coulomb
branch unlifted. For each solution there are $3$ semiclassical
limits.
\begin{enumerate}
\item $\Lambda\rightarrow 0$ and $a$ fixed and finite. Then
$m\rightarrow \pm a$,  $P_4(x)\rightarrow (x+a)^2(x-a)^2$. For
each solution the gauge group breaks into $U(4)\rightarrow
U(2)\times \widehat{U(2)}$.

\item  $\Lambda\rightarrow 0$, $a\sim \Lambda^p$ with $p\neq 1$. Then
$m\rightarrow 0$, and we can keep $v=\frac{\Lambda^3}{a^2}$ fixed. The
two solutions become $P_4(x)\rightarrow x^3(x\pm v)$. For each
solution the gauge group breaks into $U(4)\rightarrow U(1)\times
\widehat{U(3)}$.

\item $\Lambda\rightarrow 0$, $a\sim \Lambda$. Then it is possible
to keep $m$ finite by taking the limit $(\frac{a}{\Lambda}-1)\sim
\Lambda^2$. In this limit $P_4(x)\rightarrow x^3(x+m)$,
$U(4)\rightarrow \widehat{U(1)}\times U(3)$.
\end{enumerate}

Physically, how can we understand these smooth transitions? 
We will see that in all cases the exteme low
energy effective theory is just two decoupled $U(1)$s. 
For the breaking
pattern $U(2)\times \widehat{U(2)}$, the $U(2)$ factor
without flavors  is confined and leaves with one decoupled $U(1)$. 
The $U(2)$ factor with flavors is in the $r=0$ branch; so the $SU(N-r)=SU(2)$ part 
is confined and there is a decoupled $U(1)$ left.  All told we have a $U(1) \times U(1)$ at low energies. For the breaking pattern $\widehat{U(3)}\times U(1)$, the
$SU(3) \subset U(3)$ is confined in the $r=0$ branch and $U(1)\subset U(3)$ is
decoupled.  Again we find a low energy $U(1) \times U(1)$.
Finally, for the breaking pattern $U(3)\times \widehat{U(1)}$,
$SU(3)\subset U(3)$ is confined. 
Thus we have a $U(1) \times U(1)$ at low energies in this case also.\footnote{To show that the full low energy physics is identical in all these cases we should also really discuss what happens to the charged fields, but we will not do that here.}
The fact that the low energy physics is the same in all the semiclassical limits is a necessary, but not sufficient, condition for the existence of smooth transitions.

To count the number of vacua, we first need  fix $m$ and
$\alpha$ and find the number of solutions for $a$ and the shifted
constant $b$. Replacing $x\rightarrow (x+b)$ and $m\rightarrow
m-b$ in (\ref{W42}) we obtain 
\bea
W'(x) & = & (x+m)(x+\alpha)=( (x+b)+(m-b))( (x+b)+(\alpha-b)) \nonumber \\
& = & (x+b)^2 + ((m-b)+(\alpha-b))(x+b)+(m-b)(\alpha-b) \nonumber \\
& \equiv &
(x+b)^2+\frac{(m-b)\Lambda^3}{a^3}(x+b)-a^2(1+\frac{\Lambda^3}{a^3}) \label{wprime}
\eea This leads to the  equations
\begin{eqnarray}
\alpha-b+m-b&=& \frac{(m-b)\Lambda^3}{a^3} \nonumber \\
(\alpha-b)(m-b)&=&-a^2(1+\frac{\Lambda^3}{a^3}) 
\end{eqnarray}
By eliminating $b$ from these equations ($b=\frac{m+\alpha-\frac{m\Lambda^3}{a^3}}{2-\frac{\Lambda^3}{a^3}}$),
we find that $a$ must satisfy the equation
\begin{eqnarray} \label{a42}
(a^3+\Lambda^3)(2a^3-\Lambda^3)^2-a^4(a^3-\Lambda^3)(m-\alpha)^2=0
\end{eqnarray}
For fixed $m$ and $\alpha$,  we find nine solutions which have a  different
behavior when $\Lambda\rightarrow 0$. When $\Lambda=0$, we find two solutions with $a\sim\pm \frac{m-\alpha}{2}$ and seven solutions at $a=0$. The two non-zero solutions correspond to two vacua in  $U(2)\times U(2)$. We can analyze the seven solutions with $a=0$ more carefully. 
Assume $a\sim \Lambda^p$ in the $\Lambda \rightarrow 0 $ limit. Then we can have the following different cases: If $p>1$, (\ref{a42}) reduces to
$\Lambda^9+ a^4 \Lambda^3 (m-\alpha)^2=0$ which leads to  four
solutions $a\sim \Lambda^{3/2}$. These four solutions thus  correpond to four vacua in the classical limit 
 $U(1)\times \widehat{U(3)}$. If $p<1$, (\ref{a42}) reduces to $4 a^9- a^7
(m-\alpha)^2=0$ and the solution is $a\sim \Lambda^0$ which contradicts our assumption that $a\rightarrow 0$ so we should discard this soluton. If $p=1$,
 (\ref{a42}) reduces to $\Lambda^9 -a^4(a^3-\Lambda^3)(m-\alpha)^2=0$
which has a solution only when $a^3-\Lambda^3 \sim \Lambda^5$, so
$a\sim \rho(\Lambda+v \Lambda^3)$ with $\rho^3=1$. These three
solutions lead three vacua in the classical limit $\widehat{U(1)}\times U(3)$.

We can match these numbers of various solutions with the counting of vacua obtained in the semiclassical limits. 
In the  $r=0$ branch, $U(2)\times U(2)$
has four vacua. We found two vacua in the  $(2,0)/(0,2)$
distribution of double roots and  two, $(1,1)$ distribution. For $\widehat{U(1)}\times U(3)$,
we expected three vacua from
the confining $U(3)$ factor and they are indeed found in the  $(1,1)$
distribution above. For $U(1)\times \widehat{U(3)}$, we expect $2N_c-N_f=6-2=4$ vacua and they are found also
in  $(1,1)$ distribution. On the $(1,1)$ distribution of double roots,  vacua in three different classical limits smoothly interplate between each other.

\subsubsection{ $\mathbf{N_f=4}$}
In this case we can have $r=0$, $r=1$ and $r=2$ branches. The possible semiclassical limits on various $r$-th
baryonic and non-baryonic branches  can be summarized in
following table: 
\TABLE[h]
{\begin{tabular}{|c|c|c|c|c|} \hline
 & $\widehat{U(2)}\times U(2)$ & $\widehat{U(1)} \times U(3)$  & $\widehat{U(3)}\times U(1)$ & connection\\  \hline
$r=2$ & B & & N & \\ \hline
$r=1$ & N & B & N/B  &$\widehat{U(1)} \times U(3)\stackrel{B}{\leftrightarrow}  \widehat{U(3)}\times U(1)$ \\  \hline
$r=0$ & N & N & N &$\widehat{U(2)} \times U(2){\leftrightarrow}  \widehat{U(3)}\times U(1)$ \\   & & & &  
$\widehat{U(1)} \times U(3){\leftrightarrow}  \widehat{U(3)}\times U(1)$ \\  \hline
\end{tabular}
\caption{See caption of Table \ref{U4-N_f=2-table}. The symbol $\stackrel{B}{\leftrightarrow}$ indicates that the interpolation occurs on the baryonic branch.\label{U4-N_f=4-table}
}}

\paragraph{\underline{Non-baryonic $r=2$ branch:}}
By using the addition map,  this case can be reduced to a 
$U(2)$ gauge theory with no flavors in the $r=0$ branch, where we have
only one breaking  pattern $U(2)\rightarrow U(1)\times U(1)$. This leads to the unique semiclassal limit,  $U(4)\rightarrow \widehat{U(3)}\times U(1)$. There is only one vacuum in both the weak and strong coupling regions. 

\paragraph{\underline{Baryonic $r=2$ branch:}}
To obtain the semiclassical limit $U(4)\rightarrow U(2)\times \widehat{U(2)}$, we consider the baryonic branch and require an  extra double roots in the curve, which now factorizes as
\be
y^2= x^4 (x+a)^2 [ (x+a)^2- 4\eta \Lambda^2 ],~~~~or~~~P_2(x)+2\eta \Lambda^2
=(x+a)^2
\ee 
Furthermore, to find the $\CN=1$ vacua, we minimize the superpotential on this factorization locus. It is easy to see that this is essentially the same problem\footnote{This is not a coincidence. By the addition map, the $U(4) \rightarrow U(2) \times \widehat{U(2)} $ theory with four flavors charged under one of the $U(2)$ factors in the $r=2$ branch can be reduced to a $U(3) \rightarrow \widehat{U(1)} \times U(2)$ with two flavors under the $U(1)$ factor in the $r=1$ branch.} as in  (\ref{U3-Nf2-r-1}) where we found that 
$a=(\alpha\pm \sqrt{\alpha^2-4 \eta \Lambda^2})/2$. Taking only $+$ sign\footnote{The solution with the $-$ sign will not lead to two gauge group factors in the semiclassical limit. }
we have two solutions corresponding to $\eta=\pm1$. In the weak coupling region, we also have two vacua: The $U(2)$ factor with two flavors has one baryonic vacuum, while the other $U(2)$ factor will give rise to two vacua, leading to a total of two vacua in the $\widehat{U(2)} \times U(2)$ theory on this branch.

\paragraph{\underline{Non-baryonic $r=1$ branch:}}
By using the addition map, we can reduce this case to
a $U(3)$ with two flavors in the  $r=0$ branch which we analyzed carefully. 
We found that there are  two vacua for $U(3)\rightarrow \widehat{U(2)}\times U(1)$
and two vacua for the breaking pattern $U(3)\rightarrow \widehat{U(1)}\times U(2)$. Thus, by the addition map,  the 
$U(4)$ would have two vacua for $U(4)\rightarrow \widehat{U(3)}\times U(1)$ and  two vacua for $U(4)\rightarrow \widehat{U(2)}\times U(2)$.

\paragraph{\underline{Baryonic $r=1$ branch:}}
Via the addition map, this reduces to the baryonic $r=0$ branch of a $U(3)$ theory with two
flavors which we analyzed in Sec. \ref{sec:u3}. The two semiclassical limits: $U(4)\rightarrow \widehat{U(1)}\times U(3)$ and  $U(4)\rightarrow \widehat{U(3)}\times U(1)$ can be shown to be continuously connected on this branch.

\paragraph{\underline{Non-baryonic $r=0$ branch in $(2,0)/(0,2)$ 
distributions:}} On this branch with the $(0,2)$ and $(2,0)$ distribution of roots, we have 
$P_4(x)-2 \eta \Lambda^2 (x+m)^2=(x+a_1)^2(x+a_2)^2$, then
\begin{eqnarray}
F_4(x) & = & (x+a_1)^2(x+a_2)^2+4\eta\Lambda^2
(x+m)^2\nonumber \\& = &(x^2+(a_1+a_2)x+a_1a_2+2\eta\Lambda^2)^2+{\cal O}(x)
\end{eqnarray}
So $W^{\prime}(x)=x^2+(a_1+a_2)x+a_1a_2+2\eta\Lambda^2$.  There
are two solutions: $m=\frac{1}{2}(a_1+a_2\pm
\sqrt{(a_1-a_2)^2-8\eta\Lambda^2})$. In semiclassical limit, each solution leads to the breaking pattern $U(4)\rightarrow \widehat{U(2)}\times U(2)$.

To count vacua, we fix $m,\alpha$ and solve $a_1,a_2$:
\be
m+\alpha= a_1+a_2,~~~~~~m\alpha=a_1a_2+2\eta \Lambda^2
\ee
Since $a_1$ and $a_2$ are symmetric, we have only
one solution for each $\eta=\pm 1$.

\paragraph{\underline{Non-baryonic $r=0$ branch in the $(1,1)$ 
distribution:}} By shifting $x$ by a constant, we can arrange the two double roots to be at $x=a$ and  $x=-a$. The general case can be
recovered by shifting  by a constant $b$: $x\rightarrow x+b$, $m\rightarrow
m-b$. The factorization we need is
\begin{eqnarray}
P_4(x)+2\Lambda^2(x+m)^2&= &(x-a)^2\Bigl((x+a)^2+\frac{\Lambda^2}{a^3}((m^2-a^2)x+2ma(m-a))\Bigr)
\nonumber \\
P_4(x)-2\Lambda^2(x+m)^2 &= &(x+a)^2\Bigl((x-a)^2+\frac{\Lambda^2}{a^3}((m^2-a^2)x+2ma(-m-a))\Bigr) \nonumber
\end{eqnarray}
and we find
\begin{eqnarray}
F_4(x) &=& \Bigl((x+a)^2+\frac{\Lambda^2}{a^3}((m^2-a^2)x+2ma(m-a))\Bigr)
\Bigl((x-a)^2 \nonumber \\ && +\frac{\Lambda^2}{a^3}((m^2-a^2)x+2ma(-m-a))\Bigr) \nonumber \\
&=&
\Bigl(x^2+\frac{(m^2-a^2)\Lambda^2}{a^3}x-a^2(1+\frac{2m\Lambda^2}{a^3})\Bigr)^2+{\cal
O}(x)
\end{eqnarray}
So $W^{\prime}(x)$ is given by 
\begin{equation} \label{W44}
W^{\prime}(x)=x^2+\frac{(m^2-a^2)\Lambda^2}{a^3}x-a^2(1+\frac{2m\Lambda^2}{a^3})
\end{equation}
$m$ has to satisfy the following equation:
 \begin{equation} \label{17}
m^3-\frac{a^3}{\Lambda^2}m^2+a^2m+\frac{a^5}{\Lambda^2}=0
\end{equation}
This has three solutions which we denote by $m_1$, $m_2$ and $m_3$.  Notice that
 (\ref{17}) has the symmetry $m\rightarrow -m$ and $a\rightarrow -a$.
We now consider the different semiclassical limits.
\begin{enumerate}
\item $\Lambda\rightarrow 0$ and $a$ fixed and finite.
Then  from (\ref{17}),  find two solutions 
$m_1=a$ or $m_2=-a$ when  the second and fourth term in
(\ref{17}) dominates, and the third solution $m_3\sim
\frac{a^3}{\Lambda^2}$ blows up in the $\Lambda \rightarrow 0$ limit . We will ignore this solution. For $m_1$ and $m_2$ we obtain $P_4(x)\rightarrow (x+a)^2(x-a)^2$ which implies that
the breaking pattern is  $U(4)\rightarrow \widehat{U(2)}\times U(2)$.

\item  $\Lambda\rightarrow 0$, and $a\sim \Lambda^p$ with
$0<p\leq 1$. The asymptotic behavior of $m_{1,2,3}$ can be read
off from (\ref{17}). We find the solutions  $m_{1,2}\sim \pm a$ when
second and fourth term in (\ref{17}) dominate and $m_3\sim
\frac{a^3}{\Lambda^2}$ when the first and second terms dominate. 
Thus for $m_{1,2}$ we obtain $P_4(x)\rightarrow x^4$, which yields a 
singular limit since there is only one gauge group factor. For $m_3$ if
$0<p<\frac{2}{3}$ the solution blows up and should be discarded. If
$\frac{2}{3}<p\leq 1$ we obtain $P_4(x)\rightarrow x^4$, which is a
singular limit. 
For $p=\frac{2}{3}$
we obtain a smooth semiclassical limit. In this case
$m_3=\frac{a^3}{\Lambda^2}$, and we obtain $P_4(x)\rightarrow
x^3(x+m_3)$. Hence the breaking pattern is $U(4)\rightarrow
\widehat{U(1)}\times U(3)$.

\item  $\Lambda\rightarrow 0$, and $a\sim \Lambda^p$ with $p>1$.
The asymptotic behavior of $m_{1,2,3}$ can be again read off from
(\ref{17}). We find $m_{1,2}\sim \pm ia$ and $m_3\sim -\frac{a^3}{\Lambda^2}$ 
For $m_{1,2}$ we get $P_4(x)\rightarrow x^3(x-\frac{2\Lambda^2}{a})$,
which is a singular limit unless $p=2$, in which case the gauge
group breaks into $U(4)\rightarrow U(1)\times \widehat{U(3)}$. For $m_{3}$ we have $P_4(x)\rightarrow x^3(x-\frac{\Lambda^2}{a})$, which is a singular limit unless
$p=2$, in which case the gauge group breaks into $U(4)\rightarrow
U(1)\times \widehat{U(3)}$.
\end{enumerate}
Something new has happened here. To determine classical limits, we take a point on the factorization locus (parameterized by $a$ and $m$) and then determine the superpotential which would yield that point as its minimum. That leads to the consistency condition (\ref{17}), which has three solutions $m_1, m_2$ and $m_3$.  The three solutions lead to different branches and different interpolation patterns:
$m_{1,2}$ smoothly interpolate between $U(4)\rightarrow \widehat{U(2)}\times U(2)$  and  $U(4)\rightarrow
U(1)\times \widehat{U(3)}$.   $m_3$ smoothly interpolates between
 $U(4)\rightarrow \widehat{U(1)}\times U(3)$ and
$U(4)\rightarrow U(1)\times \widehat{U(3)}$. We have not encountered this phenomenon in our earlier examples. For example, in the $U(4)$ theory with two flavors, 
the three classical limits  smoothly inteplate between each other with the same choice of $m$. Notice that for the $U(4)$ theory with four flavors, the limit $\widehat{U(2)} \times U(2)$ is not smoothly connected with $\widehat{U(1)} \times U(3)$. 

To count the number of vacua, we need to first fix $m$ and
$\alpha$ and find the number of solutions for $a$ and the shifted
constant $b$. From (\ref{W44})  we obtain the equations
\begin{eqnarray}
\alpha-b+m-b&=& \frac{((m-b)^2-a^2)\Lambda^2}{a^3} \nonumber \\
(\alpha-b)(m-b)&=&-a^2(1+\frac{2(m-b)\Lambda^2}{a^3})
\end{eqnarray}
Eliminating 
$b=\frac{a^4(m+\alpha)+2a^3\Lambda^2+2m\Lambda^4-ma\Lambda^2(m-\alpha)}
{2a^4+2\Lambda^4-a\Lambda^2(m-\alpha)}$, we obtain the following equation for
$a$:
\begin{eqnarray}
&&\hspace{-0.5in}4a^8-(m-\alpha)^2a^6+4\Lambda^4a^4+\Lambda^2(m-\alpha)^3a^3
-5\Lambda^4(m-\alpha)^2a^2 \nonumber \\
& & ~~~~~~~~~~~~~~~~~~~~~~~~~~~~~~~~~~~~~+8\Lambda^6(m-\alpha)a-4\Lambda^8=0  \label{a44}
\end{eqnarray}
Thus $a$ has eight solution.  We
keep $m$ and $\alpha$ fixed and find the asymptotic behavior of
the eight roots when $\Lambda\rightarrow 0$. First by setting
$\Lambda=0$ in (\ref{a44}) we find two roots at $a\sim\pm
\frac{m-\alpha}{2}$ and six others $a\rightarrow 0$. 
The two non-zero solutions (which correspond to $m_{1,2}$ in case 1 above) lead to a semiclassical limit $U(4)\rightarrow U(2)\times \widehat{U(2)}$ . 
We now analyze the six solutions for $a$ which vanish in the $\Lambda\rightarrow 0 $ limit. 
Assume
$a\sim \Lambda^p$.  Then from (\ref{a44}) we find that 
$a\sim \Lambda^{\frac{2}{3}}$ or $a\sim \Lambda^2$. For $a\sim
\Lambda^{\frac{2}{3}}$ the dominant terms in (\ref{a44}) give
$-(m-\alpha)^2a^6+\Lambda^2(m-\alpha)^3a^3=0$ and  we find three roots
with $a^3\sim (m-\alpha)\Lambda^2$. These solutions (which correspond to  $m_3$ in case (2) above) lead to the semiclassical limit
$U(4)\rightarrow \widehat{U(1)}\times U(3)$.
 For $a\sim \Lambda^2$ the dominant terms (\ref{a44}) give
\begin{equation}
\Lambda^2(m-\alpha)^3a^3-5\Lambda^4(m-\alpha)^2a^2+8\Lambda^6(m-\alpha)a-4\Lambda^8=0
\end{equation}
This implies that
$((m-\alpha)\frac{a}{\Lambda^2}-1)((m-\alpha)\frac{a}{\Lambda^2}-2)^2=0$.
We obtain one solution (corresponding to $m_3$ in case 3 above) with  $a\sim \frac{\Lambda^2}{m-\alpha}$ and two solutions (corresponding to $m_{1,2}$ in case 3) with 
$a\sim \frac{2\Lambda^2}{m-\alpha}$. 
These three solutions lead to the semiclassical limit  $U(4)\rightarrow
U(1)\times \widehat{U(3)}$. 

We now match the number of these strong coupling vacua with the number obtained in the weak coupling region. 
In $r=0$ branch,  
$\widehat{U(2)}\times U(2)$ has four vacua where two from confining $U(2)$ factor 
and two from the $U(2)$ factor with four flavors (since $N_f>N_c$, we have
$N_c$ vacua). Two of these four vacua are in $(2,0)/(0,2)$
distribution and two, $(1,1)$ distribution. $\widehat{U(1)}\times U(3)$ has three vacua in the  $(1,1)$ distribution while
 $U(1)\times \widehat{U(3)}$ has three vacua (equal to $N_c=3$) in  $(1,1)$ distribution.

\subsubsection{$\mathbf{N_f=6}$}
In this case we can have $r=0$, $r=1$,$r=2$ and $r=3$ branches. The possible
baryonic and non-baryonic branches can be summarized in Table \ref{U4-N_f=6-table}.
\TABLE[h]
{\begin{tabular}{|c|c|c|c|c|} \hline
 & $\widehat{U(2)}\times U(2)$ & $\widehat{U(1)} \times U(3)$  & $\widehat{U(3)}\times U(1)$ & connection\\  \hline
$r=3$ &    &  & B& \\ \hline
$r=2$ & B & & N & no\\ \hline
$r=1$ & N & B & N  &no\\  \hline
$r=0$ & N & N & N &no \\   \hline
\end{tabular}
\caption{Summary of $U(4)$ with six flavors. \label{U4-N_f=6-table}
}}

\paragraph{\underline{Baryonic $r=3$ branch:}}
Via the addition map, we can reduce this to a  $U(2)$ theory with two flavors in the  $r=1$ branch. 
 Thus we conclude that on this branch, the $U(4)$ theory with six flavors has a unique classical limit
with breaking pattern  $U(4)\rightarrow \widehat{U(3)}\times U(1)$.

\paragraph{\underline{Baryonic $r=2$ branch:}}
Via the addition map, we can reduce this
to the barynoic $r=1$ branch of $U(3)$ with four
flavors. There are two vacua coming from the confining 
$U(2)$ factor in the classical limit $U(4)\rightarrow U(2)\times
\widehat{U(2)}$.

\paragraph{\underline{Non-baryonic $r=2$ branch:}}
By using the addition map, we can reduce it to a  $U(2)$ with two flavors in the $r=0$. For this theory, we found only one solution which leads to the classical limit $U(2)\rightarrow U(1)\times \widehat{U(1)}$.Thus we conclude that the $U(4)$ theory with six flavors has a classical limit with the breaking pattern  $U(4)\rightarrow \widehat{U(3)} \times U(1)$. 
The SW curve factorizes as follows
\be
y^2=x^4[(x+a)^2(x+b)^2-4\Lambda^2 x^2]
\ee
Extremizing the  superpotential, we get one solution $(a,b)=(0,\alpha)$, which will lead to the  curve $y^2=x^6[(x+\alpha)^2-4\Lambda^2]$. This is the form of the curve in
an  $r=3$ branch. We have seen this phenomenon before: this non-baryonic branch actually lies inside the baryonic $r=3$ branch.

\paragraph{\underline{Non-baryonic $r=1$ branch:}}
By the addition map, we can reduce this  to a $U(3)$ theory with four flavors in the  $r=0$ branch.  In this reduced theory,  we found
one vacuum which in the semiclassical limit became a vacuum of $\widehat{U(2)}\times U(1)$  and two vacua which 
in the semiclassical limit became vacua of  $\widehat{U(1)}\times U(2)$.
Going back to $U(4)$, we conclude that there is one vacuum which leads to 
 $U(4)\rightarrow \widehat{U(3)} \times U(1)$ and two vacua for $U(4)\rightarrow \widehat{U(2)} \times U(2)$.

\paragraph{\underline{Baryonic $r=1$ branch:}}
After factoring out  $(x+m)$ factor from $P_4(x)$ we require
\begin{eqnarray}
P_3(x)-2\Lambda^2(x+m)=(x+a+b)^2
(x+a-2\Lambda-2b-\frac{2(m-a)\Lambda}{b}) \nonumber \\
P_3(x)+2\Lambda^2(x+m)=(x+a-b)^2 (x+a+2\Lambda+2b-\frac{2
(m-a)\Lambda}{b})
\end{eqnarray}
with the constraint $b^2(b+\Lambda)=(m-a)^2\Lambda$. In
semiclassical limit $\Lambda\rightarrow 0$ we
find  $b\rightarrow 0$. Also  $(\frac{(m-a)\Lambda}{b})^{2}=\Lambda
(b+\Lambda)\rightarrow 0$, so $P_3(x)\rightarrow (x+a)^3$.
There are three vacua for a fixed superpotential.

\paragraph{\underline{Non-baryonic $r=0$ branch in $(2,0)/(0,2)$ 
distributions:}} On this branch with $(0,2)$ and $(2,0)$ distribution of roots, we have
$P_4(x)-2\eta\Lambda (x+m)^3=(x+a_1)^2(x+a_2)^2$. Then (see (\ref{flavor-F_2n}))

\begin{eqnarray}
& & F(x)+\frac{4\Lambda^2(x+m)^6}{(x+a_1)^2(x+a_2)^2}\nonumber \\
&&~~~~~~ =(x^2+(a_1+a_2)x+a_1a_2+2\eta\Lambda(x-a_1-a_2+3m))^2+{\cal
O}(x)
\end{eqnarray}
and
$W^{\prime}(x)=x^2+(a_1+a_2)x+a_1a_2+2\eta\Lambda(x-a_1-a_2+3m)$.
We find two solutions: $m=\frac{1}{2}(a_1+a_2-4\eta\Lambda\pm
\sqrt{(a_1-a_2)^2+16\Lambda^2})$.  In semiclassical limit, each solution leads to
$U(4)\rightarrow \widehat{U(2)}\times U(2)$.

To count the number of vacua, we fix $m$ and $\alpha$ and solve
$a_1$ and $a_2$. Since $a_1$ and $a_2$ are symmetric, we have only
one solution for each $\eta=\pm 1$.

\paragraph{\underline{Non-baryonic $r=0$ branch in the $(1,1)$ 
distribution:}} By a shift in  $x$, we can arrange the two double roots to be at $x=a$ and  $x=-a$. The general case can be recovered  after shifting by  a constant $b$: $x\rightarrow x+b$, $m\rightarrow
m-b$. The factorization we need is
\begin{eqnarray}
P_4(x)+2\Lambda (x+m)^3=(x-a)^2z^+
\nonumber \\
P_4(x)-2\Lambda (x+m)^3=(x+a)^2z^-
\label{whatever}
\end{eqnarray}
where $z^+$ and $z^-$ are
\begin{eqnarray}
z^+=(x+a)^2+\frac{\Lambda}{a^3}((m^3+2a^3-3ma^2)x+a(a^3+2m^3-3m^2a))
\nonumber \\
z^-=(x-a)^2+\frac{\Lambda}{a^3}((m^3-2a^3-3ma^2)x+a(a^3-2m^3-3m^2a))
\end{eqnarray}
Thus we find
\begin{eqnarray}
& & F_4(x)+\frac{4\Lambda^2(x+m)^6}{(x+a)^2(x-a)^2} \nonumber \\
&=&
(x^2+\frac{m(m^2-3a^2)\Lambda}{a^3}x-\frac{a^3+3a^2\Lambda+3m^2\Lambda}{a})^2+{\cal
O}(x)
\end{eqnarray}
and hence $W^{\prime}(x)$ is given by
\begin{equation} \label{W46}
W^{\prime}(x)=x^2+\frac{m(m^2-3a^2)\Lambda}{a^3}x-\frac{a^3+3a^2\Lambda+3m^2\Lambda}{a}
\end{equation}

$m$ must satisfy the following equation:
\begin{equation}
\frac{\Lambda m^4}{a^3}-m^2+a^2+3\Lambda a=0
\end{equation}
which yields four solutions which are given by 
\begin{eqnarray}
m_{1,2}=\pm(\frac{a^3}{2\Lambda}(1+\sqrt{(1-{6\Lambda \over
a})(1+{2\Lambda \over a})}~~))^{\frac{1}{2}}
\nonumber \\
m_{3,4}=\pm(\frac{a^3}{2\Lambda}(1-\sqrt{(1-{6\Lambda \over
a})(1+{2\Lambda \over a})}~~))^{\frac{1}{2}}
\end{eqnarray}
We now discuss different semiclassical limits:
\begin{enumerate}
\item $\Lambda\rightarrow 0$ with $a$ fixed and finite. Then
$m_{1,2}$ blow up and should be discarded. On the other hand
$m_{3,4}\rightarrow \pm a$, $P_4(x)\rightarrow (x+a)^2(x-a)^2$,
and the breaking pattern is  $U(4)\rightarrow \widehat{U(2)}\times U(2)$.

\item  $\Lambda\rightarrow 0$, $a\sim \Lambda^p$ with $0<p< 1$.
In this case $\Lambda/a$ is small. Then $m_{1,2}\sim\pm
(\frac{a^3}{\Lambda})^{\frac{1}{2}}$ and $m_{3,4}\sim\pm a$. For
the solutions $m_{1,2}$, we find that $m \gg a$ in the $\Lambda \rightarrow 0$ limit. To get a  finite $z^+$, we require
$\Lambda/a^3\sim 1$,  so the constant term (in $z_+$) drops out and we get
$P_4(x)\rightarrow x^3(x+m)$ with $p=\frac{1}{3}$. In this  case
the breaking pattern is $U(4)\rightarrow \widehat{U(1)}\times U(3)$. 
From the solutions $m_{3,4}$ we always obtain a singular limit $P_4(x)\rightarrow x^4 $.

\item $\Lambda\rightarrow 0$, $a\sim \Lambda^p$ with $p\geq 1$. Then
$m_{1,2,3,4}\sim (-3a^4)^{\frac{1}{4}}$, so $P_4(x)\rightarrow
x^4$. This is always a singular limit.
\end{enumerate}

Thus we find that for the solutions $m_{1,2}$ the only smooth semiclassical limit is
$U(4)\rightarrow \widehat{U(1)}\times U(3)$. For $m_{3,4}$ the
only smooth semiclassical limit is $U(4)\rightarrow U(2)\times
\widehat{U(2)}$.  Since these two classical limits arise from different
solutions, we do not expect them to be connected to each other
smoothly. This result is quite different from the  the $U(4)$ theory with two flavors
where the three possible classical limits were all  connected through the same branch. When we increased the number of flavors to four, the three possible classical limits were connected in pairs on two different branches. With six flavors, we find no smooth interpolations between the different classical limits.

To count the number of vacua, we need to fix $m$ and
$\alpha$ and find the number of solutions for $a$ and the shifted
constant $b$. From (\ref{W46}), we obtain the following equation:
\begin{eqnarray}
\alpha-b+m-b&=& \frac{(m-b)((m-b)^2-3a^2)\Lambda}{a^3} \nonumber \\
(\alpha-b)(m-b)&=&-\frac{a^3+3a^2\Lambda+3(m-b)^2\Lambda}{a}
\end{eqnarray}
We eliminate $b=\frac{(m+\alpha)a^3+6a(\alpha+6m)\Lambda^2
+a^2(5\alpha+11m)\Lambda+m(36\Lambda^2-(m-\alpha)^2)\Lambda}{2a^3+16\Lambda
a^2-(m-\alpha)^2\Lambda+42a\Lambda^2+36\Lambda^3}$ and find the
equation for $a$:
\begin{eqnarray} \label{a46}
4a^5+52\Lambda
a^4+(268\Lambda^2-(m-\alpha)^2)a^3+(648\Lambda^3-11\Lambda(m-\alpha)^2)a^2
\nonumber \\ +36\Lambda^2(24\Lambda^2-(m-\alpha)^2)a
+\Lambda(432\Lambda^4-36\Lambda^2(m-\alpha)^2+(m-\alpha)^4)=0
\end{eqnarray}
This is a degree five polynomial for $a$,  and thus has five
roots. Keeping $m$ and $\alpha$ fixed, we find the asymptotic
behavior of the five roots when $\Lambda\rightarrow 0$. First by
setting $\Lambda=0$ in (\ref{a46}) we find two roots at $a\sim\pm
\frac{m-\alpha}{2}$ and three others with $a\rightarrow 0$. The two non-zero
solutions are $m_{3,4}$ in case (1) for which the semiclassical limit
leads $U(4)\rightarrow U(2)\times \widehat{U(2)}$. We can analyze the three vanishing solutions more carefully.  Assume that  $a\sim \Lambda^p$. Then from
(\ref{a46}), we can conclude that  $a\sim
\Lambda^{\frac{1}{3}}$ . Then the dominant terms in (\ref{a46})
give  $-(m-\alpha)^2a^3+(m-\alpha)^4\Lambda=0$, and we find three roots
$a^3\sim(m-\alpha)^2\Lambda$. These three solutions correspond to $m_{1,2}$
in case (2) above, for which   the semiclassical limit leads to  $U(4)\rightarrow
\widehat{U(1)}\times U(3)$.

We can match the number of these strong coupling vacua with the number obtained in the weak coupling region. 
 $U(2)\times \widehat{U(2)}$ has four vacua,
 two of which are in the  $(2,0)/(0,2)$ distribution and two are in the  $(1,1)$
distribution. $\widehat{U(1)} \times U(3)$ has three vacua which lie in the  $(1,1)$ distribution.

In our discussion above, we have not found the semiclassical limit $\widehat{U(3)} \times U(1)$ in the $r=0$ branch. This is because of the following reason. In (\ref{whatever}), we have implicitly assumes $a \neq 0$. When $a=0$, the SW curve is of the same form as on the $r=2 $ branch. In fact, after the minimization of the superpotential, the curve is as it is on the $r=3$ branch. In the semiclassical limit, this leads to the breaking pattern $\widehat{U(3)} \times U(1)$.\footnote{Recall that a similar phenomenon occurred in $U(3) \rightarrow U(1) \times \widehat{U(2)} $ with four flavors charged under the $U(2)$ factor. }

\subsubsection{\textbf{Can we understand this phase structure from a weak coupling point of view?}}
In our investigations of  vacua on various branches of a $U(4)$ theory with two, four and six flavors, we encountered a rich phase structure which depended significantly on the number of flavors. Can we understand the structure from the weak coupling point of view?

Recall that in Sec. \ref{weak}, we discussed the vacuum structure of an $\CN=1$  $U(N_c)$ theory with $N_f$ massless flavors in the presence of a quartic superpotential of the form $(Q\tilde{Q})^2$. We found that vacua of this theory are characterized by an index $r$ which determines the form of the meson VEV. Furthermore, for $N_f \leq N_c$, we found $2N_c-N_f$ vacua for $r < N_f/2$ ($N_c -N_f/2$ for r=$N_f/2$). For $N_f > N_c$, we found two different sets of vacua. The first set (we will call it $NB$) had $2N_c-N_f$ vacua, while the second set (we will call it $B$) had $N_f -N_c$ vacua. 
In fact the second set is located inside the baryonic branch with 
$r=N_f-N_c$. We have encountered this in the  $U(3)$ gauge theory with
four flavors and quadratic superpotential.

For $U(4)$ with $N_f$ flavors,  we can have three semiclassical limits: 
(i) $U(2) \times \widehat{U(2)}$,  (ii) $U(1) \times \widehat{U(3)}$ and (iii) $U(3) \times \widehat{U(1)}$.
Consider the case $N_f=2$. Then, in the semiclassical limit (i) and (ii), the 
  $U(2)$ and $U(3)$ factors have $N_f\leq N_c$, so we expect these factors to have $2N_c-N_f$ vacua, all belonging to the set $NB$. On the $r=0$ branch, in the classical limit (iii), the flavors are massive quantum mechanically, and that vacuum should also be considered to be part of the set $NB$. Since all three classical limits have vacua in the set $NB$, it
is possible  that these vacua are smoothly connected. 

For $N_f=4$,  in the classical limit (i), the  $U(2)$ factor has no vacuum in the set $NB$ and has two vacua in the set $B$. In the classical limit (ii), there are three vacua, two of which lie in the set $NB$ and one lies in the set $B$. In the classical limit (iii), the $U(1)$ factor has one vacuum in the set $NB$. Thus we expect vacua of (i) to be smoothly connected to vacua of (ii) (and {\em not} vacua of (iii)), while vacua of (iii) can also only connect with vacua of (ii). This is indeed the structure we found in the previous subsection.

For $N_f=6$, in the classical limit (i), the  $U(2)$ factor only has vacua in set B while  the  $U(1)$ factor in classical limit (iii) only has vacua in the set $NB$. In the classical limit (ii) the $U(3)$ factor also has vacua only in the set $B$. But for this special case, the $r=0$ branch is actually embedded inside an $r=3$ branch. Thus we expect no smooth interpolations to be possible between vacua in different classical limits, which is indeed what we found in the previous subsection. 

\section{Conclusion}
\label{conclusion}
In this paper, we have studied the phase structure of $\CN = 1$ supersymmetric $U(N)$ gauge theories with fundamental matter that arise as deformations of $\CN = 2$ SQCD by the addition of a superpotential for the adjoint chiral multiplet. The various $\CN=1$ vacua lie on different branches, some of which have multiple classical limits in which the vacua are interpreted as those of a product group theory with flavors charged under various group factors. On a given branch, the vacua are all in the same phase. What order parameters (or indices) distinguish between vacua on different branches? An obvious characterization of a branch is the global symmetry group (which in this case will be a flavor symmetry). 
We have mentioned other several quantities, which provide necessary conditions for vacua to be on the same branch. 
One such index is $r$, which characterizes the meson VEV in various vacua in the weak coupling region, and labels the root of the $r$-th Higgs branch in the strong coupling region.
The global symmetry group must be the same on each branch. On the  $r$-th branch, the global flavor symmetry is broken as $SU(N_f)\rightarrow SU(N_f-2r)\times U(1)^r$. 
  Another fact which distinguishes the different branches is if they are `baryonic' or `non-baryonic'. These two types of branches differ from each other in the number of condensed monopoles, and hence have different number of $U(1)$s at low energies.  Furthermore, a finer distinction is possible on the non-baryonic $r$-th branches which arise when $N_f>N_c$ and $r < N_f - N_c$.
In these cases, there are two types of non-baryonic branch.  In the strong coupling region one arises from  a generic non-baryonic root, while the other is special case arising when the non-baryonic root lies inside the  baryonic root.   These two kinds of strong coupling vacua match up with two types weak coupling vacua in which the meson matrix is degenerate and non-degenerate.
 
We have also defined a multiplication index $t$ in our theory with flavors. In \cite{Cachazo:2002zk}, in the absence of flavors, this index is related to a ``confinement index'' which had a physical meaning in terms of representations whose Wilson loops did not show an area law. The meaning of the confinement index is not clear in the presence of massless flavors because perfect screening prevents Wislon loops in any representation from having an area law and the naive definition of the confinement index  always yields a trivial result. Perhaps the connection to a matrix model \cite{flavornati,bomassless,Roiban} can be exploited to define a more useful analog of the confinement index. In any case, the various indices that are available to us are not refined enough to provide sufficient conditions which determine the phase structure completely.

\section*{Acknowledgements}
We thank Alex Buchel, Freddy Cachazo, Josh Erlich, Yang-Hui He, Vishnu Jejjala and Nati Seiberg for useful discussions. B. F. thanks the high energy theory group at the University of Pennsylvania for their kind hospitality through the course of this work. V.B. thanks the George P. and Cynthia W. Mitchell Institute for Fundamental Physics for hospitality as this work was being completed. 
Work on this project at the University of Pennsylvania was supported
by the DOE under cooperative research agreement DE-FG02-95ER40893, and
by an NSF Focused Research Grant DMS0139799 for ``The Geometry of
Superstrings".   This research was also supported by the Institute for
Advanced Study, 
under the NSF grant PHY-0070928. 

\bibliographystyle{JHEP}

\begin{thebibliography}{99}
\bibitem{Cachazo:2002zk}
F.~Cachazo, N.~Seiberg and E.~Witten,
``Phases of N = 1 supersymmetric gauge theories and matrices,''
arXiv:hep-th/0301006.
\bibitem{tamar}
T.~Friedmann,
``On the quantum moduli space of M theory compactifications,''
Nucl.\ Phys.\ B {\bf 635}, 384 (2002)
[arXiv:hep-th/0203256].
\bibitem{ferrari}
F.~Ferrari,
``Quantum parameter space and double scaling limits in N = 1 super  Yang-Mills theory,''
arXiv:hep-th/0211069; 
F.~Ferrari,
``Quantum parameter space in super Yang-Mills. II,''
arXiv:hep-th/0301157.
\bibitem{dv1}
R.~Dijkgraaf and C.~Vafa,
``Matrix models, topological strings, and supersymmetric gauge theories,''
Nucl.\ Phys.\ B {\bf 644}, 3 (2002)
arXiv:hep-th/0206255.
\bibitem{dv2}
R.~Dijkgraaf and C.~Vafa,
``On geometry and matrix models,''
Nucl.\ Phys.\ B {\bf 644}, 21 (2002)
arXiv:hep-th/0207106.
\bibitem{dv3}
R.~Dijkgraaf and C.~Vafa,
``A perturbative window into non-perturbative physics,''
arXiv:hep-th/0208048.
\bibitem{Ahn:2003cq}
C.~Ahn and Y.~Ookouchi,
``Phases of N = 1 supersymmetric SO / Sp gauge theories via matrix
model,''
arXiv:hep-th/0302150.
\bibitem{oz}
A.~Brandhuber, H.~Ita, H.~Nieder, Y.~Oz and C.~Romelsberger,
``Chiral Rings, Superpotentials and the Vacuum Structure of N=1 Supersymmetric Gauge Theories,''
arXiv:hep-th/0303001.

\bibitem{Cachazo:2001jy}
F.~Cachazo, K.~A.~Intriligator and C.~Vafa,
``A large N duality via a geometric transition,''
Nucl.\ Phys.\ B {\bf 603}, 3 (2001)
[arXiv:hep-th/0103067].


\bibitem{Argyres:1996eh}
P.~C.~Argyres, M.~Ronen Plesser and N.~Seiberg,
``The Moduli Space of N=2 SUSY {QCD} and Duality in N=1 SUSY {QCD},''
Nucl.\ Phys.\ B {\bf 471}, 159 (1996)
[arXiv:hep-th/9603042].

\bibitem{Intriligator}
K.~A.~Intriligator and N.~Seiberg,
``Lectures on supersymmetric gauge theories and electric-magnetic  duality,''
Nucl.\ Phys.\ Proc.\ Suppl.\  {\bf 45BC}, 1 (1996)
[arXiv:hep-th/9509066].

\bibitem{Hori:1997ab}
K.~Hori, H.~Ooguri and Y.~Oz,
``Strong coupling dynamics of four-dimensional N = 1 gauge theories from  M theory fivebrane,''
Adv.\ Theor.\ Math.\ Phys.\  {\bf 1}, 1 (1998)
[arXiv:hep-th/9706082].


\bibitem{Carlino:2000uk}
G.~Carlino, K.~Konishi and H.~Murayama,
``Dynamical symmetry breaking in supersymmetric SU(n(c)) and USp(2n(c))  gauge theories,''
Nucl.\ Phys.\ B {\bf 590}, 37 (2000)
[arXiv:hep-th/0005076].


\bibitem{deBoer:1997ap}
J.~de Boer and Y.~Oz,
``Monopole condensation and confining phase of N = 1 gauge theories via  M-theory fivebrane,''
Nucl.\ Phys.\ B {\bf 511}, 155 (1998)
[arXiv:hep-th/9708044].

\bibitem{Seiberg:1994bz}
N.~Seiberg,
``Exact results on the space of vacua of four-dimensional SUSY gauge theories,''
Phys.\ Rev.\ D {\bf 49}, 6857 (1994)
[arXiv:hep-th/9402044].

\bibitem{Seiberg:1994pq}
N.~Seiberg,
``Electric - magnetic duality in supersymmetric nonAbelian gauge theories,''
Nucl.\ Phys.\ B {\bf 435}, 129 (1995)
[arXiv:hep-th/9411149].


\bibitem{ADS}
I.~Affleck, M.~Dine and N.~Seiberg,
``Dynamical Supersymmetry Breaking In Supersymmetric QCD,''
Nucl.\ Phys.\ B {\bf 241}, 493 (1984).
\bibitem{Klemm} 

A.~Klemm, W.~Lerche, S.~Yankielowicz and S.~Theisen,
``Simple singularities and N=2 supersymmetric Yang-Mills theory,''
Phys.\ Lett.\ B {\bf 344}, 169 (1995)
[arXiv:hep-th/9411048].


\bibitem{Argyres}
P.~C.~Argyres, M.~R.~Plesser and A.~D.~Shapere,
``The Coulomb phase of N=2 supersymmetric QCD,''
Phys.\ Rev.\ Lett.\  {\bf 75}, 1699 (1995)
[arXiv:hep-th/9505100].

\bibitem{Hanany}
A.~Hanany and Y.~Oz,
``On the quantum moduli space of vacua of N=2 supersymmetric SU(N(c)) gauge theories,''
Nucl.\ Phys.\ B {\bf 452}, 283 (1995)
[arXiv:hep-th/9505075].


\bibitem{Ookouchi:2002be}
Y.~Ookouchi,
``N = 1 gauge theory with flavor from fluxes,''
arXiv:hep-th/0211287.

\bibitem{Cachazo:2002pr}
F.~Cachazo and C.~Vafa,
``N = 1 and N = 2 geometry from fluxes,''
arXiv:hep-th/0206017.


\bibitem{Feng:2002gb}
B.~Feng,
``Geometric dual and matrix theory for SO/Sp gauge theories,''
arXiv:hep-th/0212010.

\bibitem{dglvz}
R.~Dijkgraaf, M.~T.~Grisaru, C.~S.~Lam, C.~Vafa and D.~Zanon,
``Perturbative computation of glueball superpotentials,''
arXiv:hep-th/0211017.

\bibitem{CDSW}
F.~Cachazo, M.~R.~Douglas, N.~Seiberg and E.~Witten,
``Chiral rings and anomalies in supersymmetric gauge theory,''
JHEP {\bf 0212}, 071 (2002)
[arXiv:hep-th/0211170].


\bibitem{janik}
Y.~Demasure and R.~A.~Janik,
``Explicit factorization of Seiberg-Witten curves with matter from random  matrix models,''
arXiv:hep-th/0212212.


\bibitem{Douglas:1995nw}
M.~R.~Douglas and S.~H.~Shenker,
``Dynamics of SU(N) supersymmetric gauge theory,''
Nucl.\ Phys.\ B {\bf 447}, 271 (1995)
[arXiv:hep-th/9503163].

\bibitem{Ferrari:2002jp}
F.~Ferrari,
``On exact superpotentials in confining vacua,''
Nucl.\ Phys.\ B {\bf 648}, 161 (2003)
[arXiv:hep-th/0210135].

\bibitem{Balasubramanian:2002tm}
V.~Balasubramanian, J.~de Boer, B.~Feng, Y.~H.~He, M.~x.~Huang, V.~Jejjala and A.~Naqvi,
``Multi-trace superpotentials vs. Matrix models,''
arXiv:hep-th/0212082.

\bibitem{Argyres:1995jj}
P.~C.~Argyres and M.~R.~Douglas,
``New phenomena in SU(3) supersymmetric gauge theory,''
Nucl.\ Phys.\ B {\bf 448}, 93 (1995)
[arXiv:hep-th/9505062].

\bibitem{Argyres:1995xn}
P.~C.~Argyres, M.~Ronen Plesser, N.~Seiberg and E.~Witten,
``New N=2 Superconformal Field Theories in Four Dimensions,''
Nucl.\ Phys.\ B {\bf 461}, 71 (1996)
[arXiv:hep-th/9511154].

\bibitem{Eguchi:1996vu}
T.~Eguchi, K.~Hori, K.~Ito and S.~K.~Yang,
``Study of $N=2$ Superconformal Field Theories in $4$ Dimensions,''
Nucl.\ Phys.\ B {\bf 471}, 430 (1996)
[arXiv:hep-th/9603002].



\bibitem{flavornati}
N.~Seiberg,
``Adding fundamental matter to 'Chiral rings and anomalies in  supersymmetric gauge theory',''
JHEP {\bf 0301}, 061 (2003)
[arXiv:hep-th/0212225].



\bibitem{bomassless}
B.~Feng,
``Note on matrix model with massless flavors,''
arXiv:hep-th/0212274.


\bibitem{Roiban}
R.~Roiban, R.~Tatar and J.~Walcher,
``Massless flavor in geometry and matrix models,''
arXiv:hep-th/0301217.


























\end{thebibliography}

\end{document}